\documentclass[preprint,amsmath,amssymb,showkeys,nofootinbib]{revtex4}

\usepackage{graphicx}
\usepackage{dcolumn}
\usepackage{bm}
\begin{document}

\title{Comparison of Fast Amplifiers for Diamond Detectors.}

\author{M.~Osipenko}
\affiliation{INFN, Sezione di Genova, 16146 Genova, Italy}
%\altaffiliation[Also at ]{Skobeltsyn Institute of Nuclear Physiscs, 119992 Moscow, Russia}
\email{osipenko@ge.infn.it}
\author{S.~Minutoli}
\affiliation{INFN, Sezione di Genova, 16146 Genova, Italy}
\author{P.~Musico}
\affiliation{INFN, Sezione di Genova, 16146 Genova, Italy}
\author{M.~Ripani}
\affiliation{INFN, Sezione di Genova, 16146 Genova, Italy}
\author{B.~Caiffi}
\affiliation{Dipartimento di Fisica, Universit\`a di Genova, 16146 Genova, Italy}
\author{A.~Balbi}
\affiliation{INFN, Sezione di Genova, 16146 Genova, Italy}
\author{G.~Ottonello}
\affiliation{INFN, Sezione di Genova, 16146 Genova, Italy}
%\email{simula@roma3.infn.it}
\author{S.~Argir\`o}
\affiliation{Universit\`a di Torino and INFN, Sezione di Torino, 10125 Torino, Italy}
\author{S.~Beol\`e}
\affiliation{Universit\`a di Torino and INFN, Sezione di Torino, 10125 Torino, Italy}
\author{N.~Amapane}
\affiliation{Universit\`a di Torino and INFN, Sezione di Torino, 10125 Torino, Italy}
\author{M.~Masera}
\affiliation{Universit\`a di Torino and INFN, Sezione di Torino, 10125 Torino, Italy}
\author{G.~Mila}
\affiliation{Universit\`a di Torino and INFN, Sezione di Torino, 10125 Torino, Italy}

\date{\today}

\begin{abstract}
The development of Chemical Vapour Deposition (CVD) diamond detectors requests for novel signal amplifiers,
capable to match the superb signal-to-noise ratio and timing response of these detectors.
Existing amplifiers are still far away from this goal
and are the dominant contributors to the overall system noise and
the main source of degradation of the energy and timing resolution.
We tested a number of commercial amplifiers designed for diamond detector readout
to identify the best solution for a particular application. 
This application
required a deposited energy threshold below 100 keV and timing
resolution of the order of 200 ps at 200 keV.
None of tested amplifiers satisfies these requirements.
The best solution to such application found to be the Cividec C6
amplifier, which allows 100 keV minimal threshold, but its
coincidence timing resolution at 200 keV is as large as 1.2 ns.
\end{abstract}

\keywords{diamond detector amplifiers, timing resolution}

\maketitle

%\linenumbers

\section{\label{sec:introduction}Introduction}
Recent progress in the growth of Chemical Vapour Deposition (CVD) diamonds is challenging readout electronics developers.
Because of its large bandgap of 5.5 eV and very low Boron and Nitrogen impurity concentrations
$<$1-5 ppb~\cite{E6} CVD diamond has a negligible intrinsic noise at room temperature.
In fact, the mean leakage current of a typical device is less than 1 pA.
Combining this with a very high carrier mobility of about $2\times 10^{-4}$c/(V/$\mu$m),
allowing for a complete charge collection in few ns (8 ns for 500 $\mu$m thick device biased at 1 V/$\mu$m),
results in an intrinsic noise charge collected within the duration of the signal of the order of $\sim$0.05 electrons.
Except for very slow, cryogenic, charge sensitive amplifiers, %~\cite{askGatti}
able to reduce the RMS noise to a few electrons, such precision is out of reach of modern fast amplifiers.
Indeed, the best broadband amplifiers have input referred noise RMS of $10^4$ electrons
corresponding to the input referred noise of 10 $\mu$V (50 $\Omega$ input impedance) in 8 ns signal.
Even conventional charge amplifiers
with 50-300 $\mu$s decay time have more than 200 electrons RMS noise.
Therefore, for diamond detector applications the dominant (by three-five orders of magnitude)
source of noise is the readout electronics.

In the present note we present a series of tests on various fast amplifiers
aimed to identify the best solution for nuclear and particle physics applications.
These applications demand a measurement of both the energy
released in the bulk of the diamond by ionizing particles and the time.
These tests were performed using various radioactive sources.

\section{\label{sec:setup}Experimental Setup}
We selected a 4.7 x 4.7mm2, 500 um thick single crystal CVD diamond detector
capacitance can be estimated according TO the plain

We selected 4.7$\times$4.7 mm$^2$, 500 $\mu$m thick single crystal CVD diamond detector
produced by Diamond Detectors Ltd~\cite{DDL}.
The crystal has two electrodes deposited on its upper and lower major surfaces,
made of few nm of DLC followed by 100 nm of Gold.
Each contact is bonded via a gold microwire onto PCB and then
to its SMA connector in such a way that both contacts can be read out
independently.
Detector capacitance can be estimated according to the plane capacitor
equation (neglecting eventual fringe field contribution):
\begin{equation}
C_D = \frac{\varepsilon_D \varepsilon_0 A}{d} = \frac{5.7 \times 8.854\times10^{-3} pF/mm \times  4.7\times4.7 mm^2 }{0.5 mm}
= 2.23 pF
\end{equation}
\noindent This result is in good agreement with the value measured by HP 4280A capacimeter.

The diamond detector was polarized at 1 V/$\mu$m by means of Ortec 710 Bias Supply.
This value remains below detector breakdown voltage (at 1.4 V/$\mu$m discharges, likely to be attributed to crystal
defects or contact disuniformity, occur)
allowing for complete charge collection~\cite{kagan}.

\begin{figure}[!h]
\begin{center}
\includegraphics[bb=1cm 4cm 14cm 27cm, angle=270, scale=0.4]{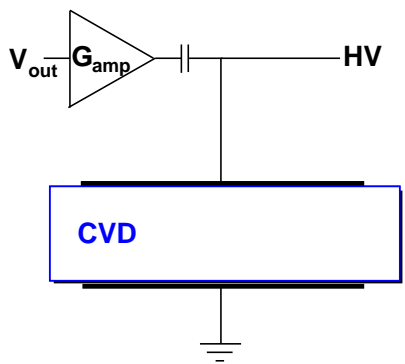}
\caption{\label{fig:setup_single} Schematic drawing of the experimental setup
in case of the one-side readout.}
\end{center}
\end{figure}

\begin{figure}[!h]
\begin{center}
\includegraphics[bb=1cm 4cm 17cm 27cm, angle=270, scale=0.4]{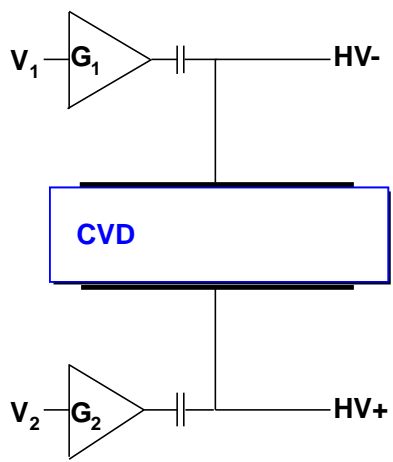}
\caption{\label{fig:setup_double} Schematic drawing of the experimental setup
in case of the two-side readout.}
\end{center}
\end{figure}

In the present work, we used two types of readout:
one-side readout, shown in Fig.~\ref{fig:setup_single}
and two-side readout, shown in Fig.~\ref{fig:setup_double}.
The one-side readout was used to characterize energy resolution of
of various amplifiers. In this case the diamond
was polarized by applying 1 V/$\mu$m bias voltage from the readout side
and the other contact was connected to ground.

The two-side readout was used to study ideal timing resolution
of the CVD diamond detector. Indeed, the major contribution
to both energy and timing resolutions come from the readout electronics
and in particular from the first amplification stage. Using the same diamond detector signal
read out independently from opposite sides by two separate amplifiers
allows to study the electronic contribution to the timing resolution
using realistic signals.

For single amplifier characterization measurements the diamond detector was connected to the
amplifier input through a 2.4 cm Huber \& Suhner SMA ``I'' connector, while the second SMA connector of the detector was shorted to ground by SMA tap.
The overall capacitance of the system in front of the amplifier (``I'' connector,
two SMA connectors of detector case and SMA tap) was found to be 6.6 pF.

The amplifier output was connected to the digitizer via 1.5 m Huber \& Suhner SucoFlex 104 cable.
The amplified signal was read out by SIS3305 digitizer with analog bandwidth of 2.2 GHz.
The digitizer was operated in four-channel interleaved 5 Gs/s mode.
For timing response measurements with small signals an additional amplifier Philips Scientific 744
with bandwidth of 1.8 GHz and gain $\times 20$ was inserted between
primary amplifier output and digitizer input.

The data measured by the digitizer was acquired through the VME bus
by Concurrent Technologies VX 813/091 Single Board Computer (SBC) which incorporates a Tsi148 VME controller.
A simple DAQ program was developed using DMA transfer of digitizer memory to the SBC memory.
The data were saved on the local Compact Flash disk and
transferred to workstation for the off-line analysis.

\section{\label{sec:firmware}Coincidence Trigger}
In order to measure coincidence events and to reduce accessible energy threshold
the standard SIS3305 digitizer firmware rev.1C.0B had to be modified.
SIS3305 digitizer features three Xilinx FPGAs on-board,
two Virtex 5 FPGAs serve as interfaces for the two 2.5 Gs/s ADC cores,
and one Virtex 4 FPGA manages the board VME interface.

Partially precompiled firmware of ADC FPGAs was acquired from the manufacturer for our purposes.
This firmware version 1C.0B included only simple internal triggers:
data acquisition is started when one ADC sample goes above or below imposed threshold in one of digitizer channels.
It also includes a Schmidt trigger, which fires when one ADC sample goes above (below)
a first threshold and turns off when after another single ADC sample is below (above) a second threshold.

The trigger was modified using the Xilinx ISE Design Suite version 13.2.

First of all a trigger that fires when a configurable number of samples is above (below)
a given threshold was implemented.
For the initialization of this parameter for each channel we used bits 30-26 of the corresponding
SIS3305\_TRIGGER\_GATE\_GT\_THRESHOLDS\_ADC1-8 and SIS3305\_TRIGGER\_GATE\_LT\_THRESHOLDS\_ADC1-8
(0x2020-0x203C and 0x3020-0x303C) registers~\cite{SIS3305}.
This allows to select the number of consecutive samples above/below threshold in selected channel
in the range from 1 to 32 (5-bit word) corresponding to time interval from 0.8 ns to 25.6 ns.
The 4.8 ns long channel's trigger-valid signal goes on when the number of consecutive samples above/below
threshold in given channel reaches the configured value.
In case of the 2 or 4 channel interleaved modes (2.5 Gs/s and 5 Gs/s, respectively)
an asynchronous AND of the corresponding channel trigger-valid signals is taken
and then synchronized with 250 MHz system clock to give the final ADC core internal trigger.
The acquisition mode is determined by the first three bits of SIS3305\_EVENT\_CONFIG\_ADC1\_4 (0x2000)
or  SIS3305\_EVENT\_CONFIG\_ADC5\_8 (0x3000) register.
In this way the real number of samples above/below threshold in 2 or 4 channel
interleaved modes is equal to the configured value multiplied by 2 or 4, respectively.
The ADC core internal trigger signal can be extended from its natural length of 4 ns
up to 128 ns by configuring bits 14-10 of SIS3305\_TRIGGER\_GATE\_GT\_THRESHOLDS\_ADC1 (0x2020) register.
This 5-bit word allows to modify the coincidence time window with precision of 4 ns.

It has to be mentioned that since many firmware blocks were provided in a precompiled form
we had to follow its existing design. In particular, the data from each ADC input channel
acquired at 1.25 Gs/s rate are split in ADC-FPGA in six parallel 208.3 MHz flows.
This architecture limits the trigger selectivity in the 2 or 4 channel
interleaved modes. Indeed, the length of trigger-valid signal from each ADC channel is 4.8 ns,
therefore imposing only one sample above/below threshold in one of interleaved modes (2 or 4 channels)
there is a probability to trigger on accidental coincidence of background hits (whose rate $R_{bkg}$) in selected channels
within $\Delta t=4.8$ ns time window of $(R_{bkg}\times\Delta t)^{2,4}$).
%Increasing the requested number of samples above/below threshold enhance the average number of samples
%overlapping in different channels leading to correct definition of ADC trigger.
In order to prevent this possible background the continuity of trigger condition
among interleaved channels can be activated by an additional configuration flag.
Setting bit-15 of SIS3305\_TRIGGER\_GATE\_GT\_THRESHOLDS\_ADC1 (0x2020) register
the difference between indexes of trigger samples within 6-sample data blocks
from interleaved ADC channels is checked to be not larger than unity.

The coincidence between two ADC cores (channels 1-4 and 5-8) is implemented in the VME-FPGA
by an asynchronous AND between two ADC core internal trigger signals, then synchronized with 125 MHz system clock
and extended to 32 ns length.
Setting bit-20 of SIS3305\_TRIGGER\_OUT\_SELECT\_REG (0x40) register
allows to route the coincidence signal onto a Lemo Trigger Out connector.
Thus connecting the latter to the Lemo Trigger In connector and enabling
the External Lemo Trigger In bit in SIS3305\_CONTROL\_STATUS (0x0) register
and External Trigger bit in SIS3305\_EVENT\_CONFIG\_ADC1\_4/5\_8 (0x2000 and 0x3000) register
allows to measure coincidence events only. The coincidence interval can be configured
with 4 ns steps up to 128 ns as explained above.

Similarly, bit-21 of SIS3305\_TRIGGER\_OUT\_SELECT\_REG (0x40) register allows
to trigger on OR of two ADC core internal triggers. This feature permits to acquire
both ADCs when one of two ADC cores had an internal trigger.

It has to be noticed that in this modified firmware the Schmidt trigger has been removed
and therefore trigger-off setting in bits 25-16 of
SIS3305\_TRIGGER\_GATE\_GT\_THRESHOLDS\_ADC1-8 and SIS3305\_TRIGGER\_GATE\_LT\_THRESHOLDS\_ADC1-8
registers (0x2020-0x203C and 0x3020-0x303C) are not more significant.
Moreover, the internal trigger logic in the 2 channel interleaved mode (2.5 Gs/s)
assumes that two pairs of channels are: 1 and 2, 3 and 4.

This modified firmware is available at~\cite{firmware2c}.

\section{\label{sec:noise}Noise Figure}
\begin{figure}[!h]
\begin{center}
\includegraphics[bb=4cm 4cm 20cm 27cm, angle=270, scale=0.4]{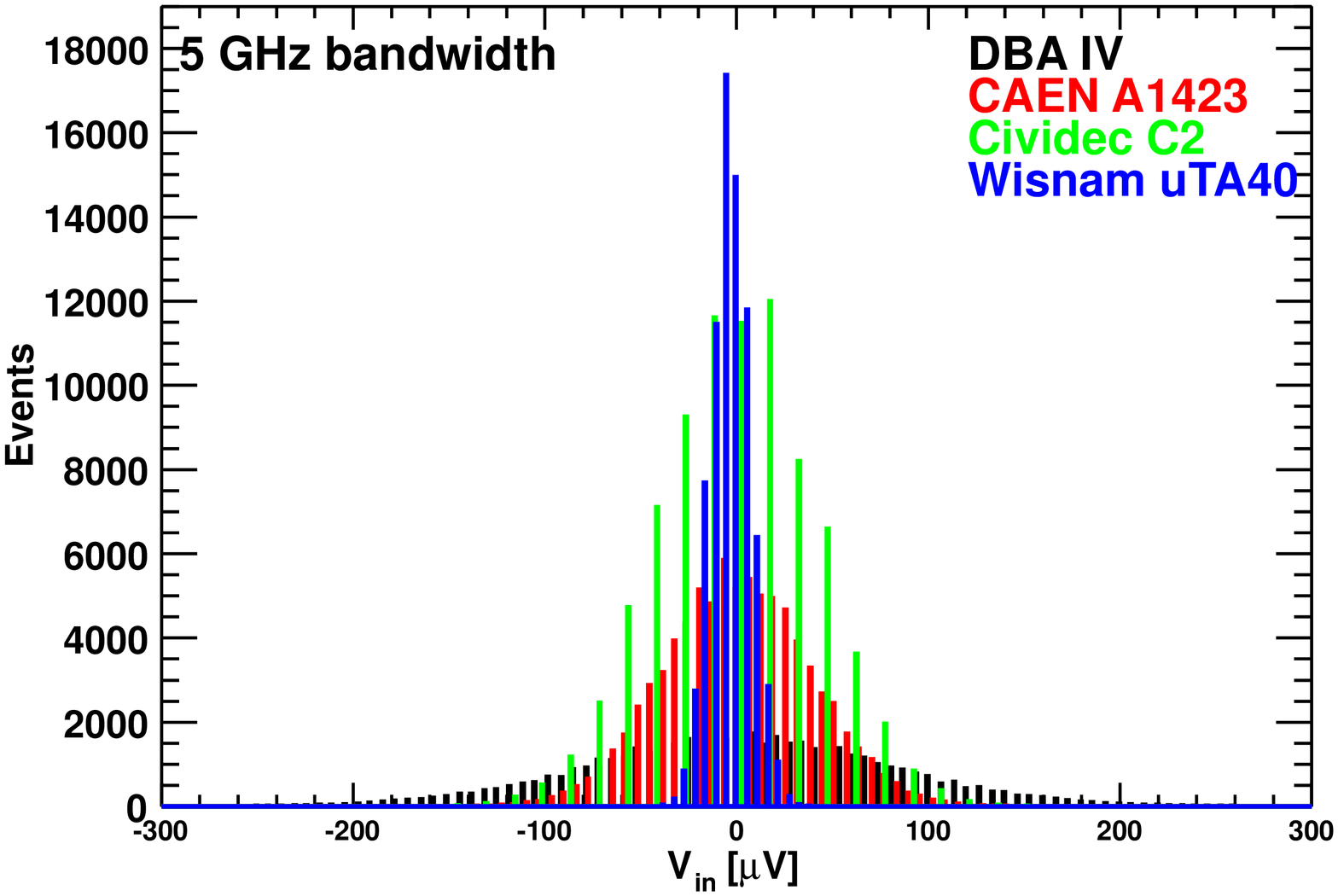}~~~%
\includegraphics[bb=4cm 4cm 20cm 27cm, angle=270, scale=0.4]{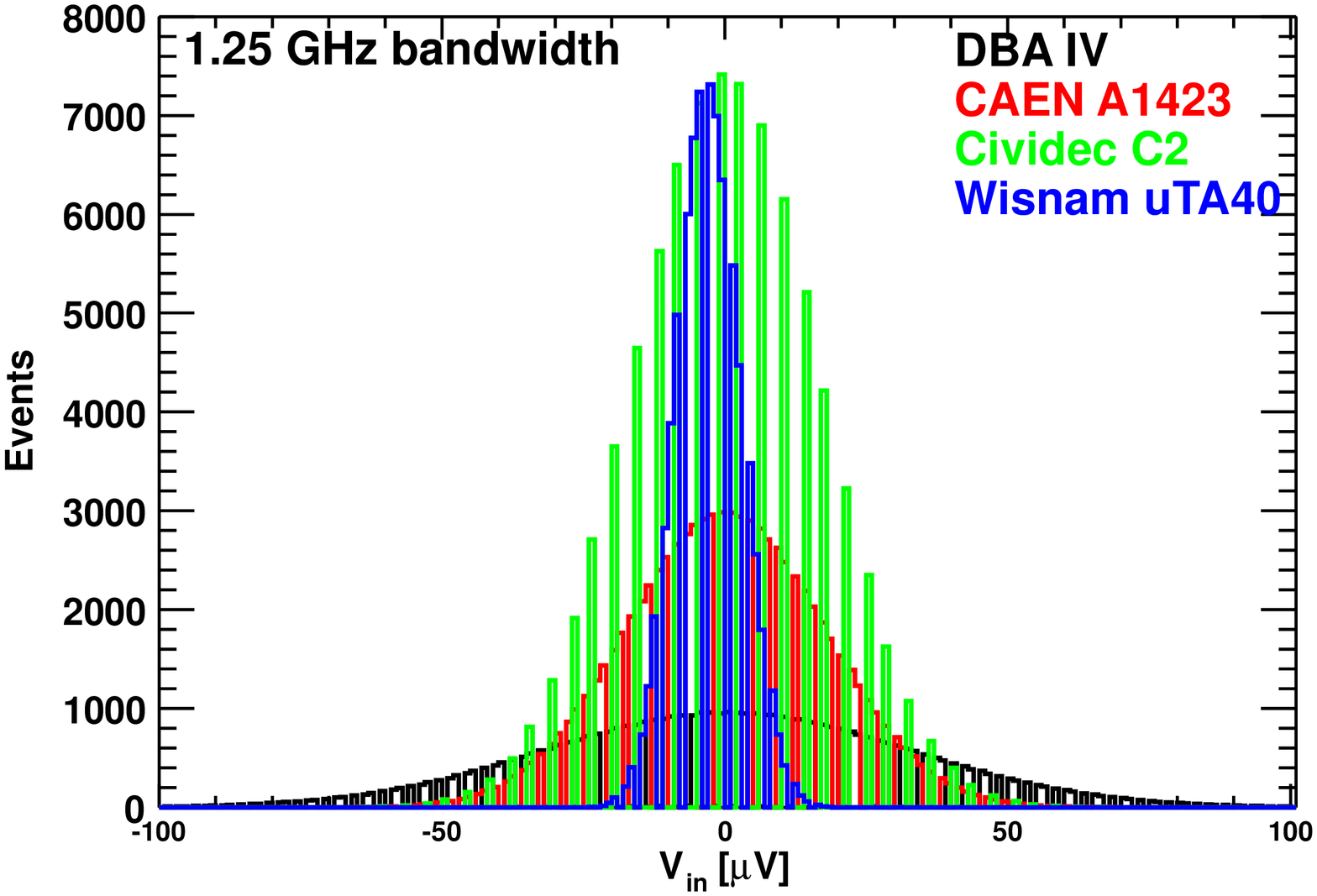}
\caption{\label{fig:noise_distr} Noise distributions as referred to input for different amplifiers
measured by SIS3305 2.2 GHz bandwidth digitizer in
5 Gs/s mode (left) and 1.25 Gs/s mode (right).}
\end{center}
\end{figure}

\begin{figure}[!h]
\begin{center}
\includegraphics[bb=4cm 4cm 20cm 27cm, angle=270, scale=0.4]{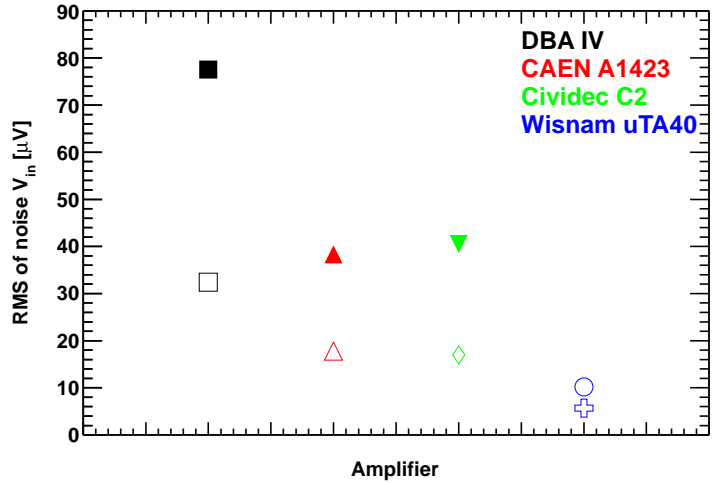}
\caption{\label{fig:noise_rms} RMS of noise as referred to input for different amplifiers.
Upper points are for 5 Gs/s mode, while lower points are for 1.25 Gs/s mode.}
\end{center}
\end{figure}

The noise level referred to the input was measured simultaneously with signal acquisition
by comparing the voltages in the samples before the signal peak to the baseline voltage.
This was done for the full digitizer sampling rate of 5 Gs/s and summing four interleaved channel
amplitudes to restrict the sampling rate to 1.25 Gs/s (SIS3305 digitizer analog bandwidth is 2.2 GHz).
The resulted distributions are shown in Fig.~\ref{fig:noise_distr} and their
RMS values are indicated in Fig.~\ref{fig:noise_rms}.

\begin{table}[!h]
\begin{center}
\caption{\label{table:noise_rms}Noise RMS values as referred to input
for different amplifiers measured with 2.2 GHz bandwidth digitizer in 5 Gs/s mode
compared to values declared by the manufacturer.}
\vspace{2mm}
\begin{tabular}{|c|c|c|} \hline
Amplifier        & $V_{noise}^{measured}$ & $V_{noise}^{declared}$ \\
                 & [$\mu$V]    & [$\mu$V]    \\ \hline
DBA IV\cite{DBA}              & 78	       & 30 (5 dB at 2 GHz) \\ \hline
CAEN A1423\cite{CAEN}         & 38	       & 26 (5 dB at 1 GHz) \\ \hline
Cividec C2\cite{Cividec}      & 41	       & 25                 \\ \hline
Wisnam $\mu$TA40\cite{Wisnam} & 10	       & NA                 \\ \hline
Cividec C6\cite{Cividec}      &  6	       &  1.2               \\ \hline
\end{tabular}
\end{center}
\end{table}

\begin{figure}[!h]
\begin{center}
\includegraphics[bb=4cm 4cm 20cm 27cm, angle=270, scale=0.4]{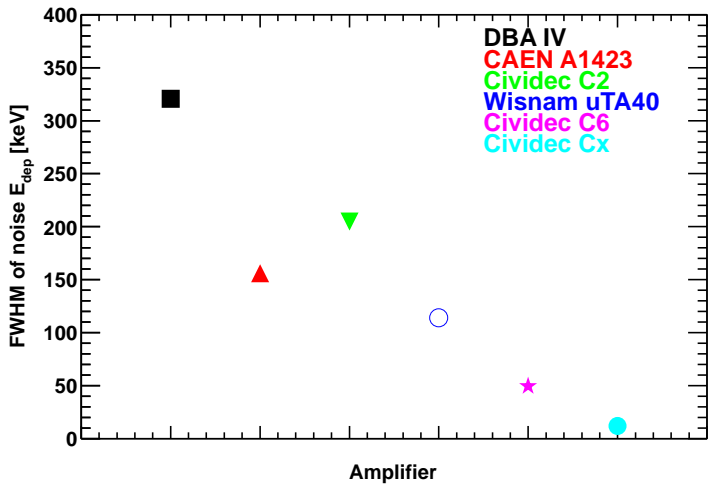}
\caption{\label{fig:edep_noise_fwhm} FWHM of noise in terms of energy deposited in diamond detector
for different amplifiers. For broadband amplifiers the signal width was taken to be 9 ns corresponding to 500 $\mu$m
single crystal CVD diamond polarized by 1 V/$\mu$m bias.}
\end{center}
\end{figure}

We converted these observed noise distributions in projected uncertainty on
the collected charge and therefore on deposited energy. To this end we integrated
noise output of the amplifier in the same time interval chosen to acquire the signal.
The width of the Gaussian distribution of such integrals was extracted.
The obtained resolution due to amplifier noise is shown in Fig.~\ref{fig:edep_noise_fwhm}.
In this graph, charge (Cividec Cx) and transimpedance (Cividec C6) amplifier resolutions are also
given for comparison. These were obtained with the same technique described so far, but
converting noise amplitude into deposited energy equivalent.

The results shown in Fig.~\ref{fig:edep_noise_fwhm} demonstrate the expected superiority of charge and transimpedance
amplifiers in terms of energy resolution. Among broadband amplifiers
the best resolution of 150 keV was achieved with CAEN A1423.

\section{\label{sec:edep_res}Energy resolution}
\begin{figure}[!h]
\begin{center}
\includegraphics[bb=4cm 4cm 20cm 27cm, angle=270, scale=0.4]{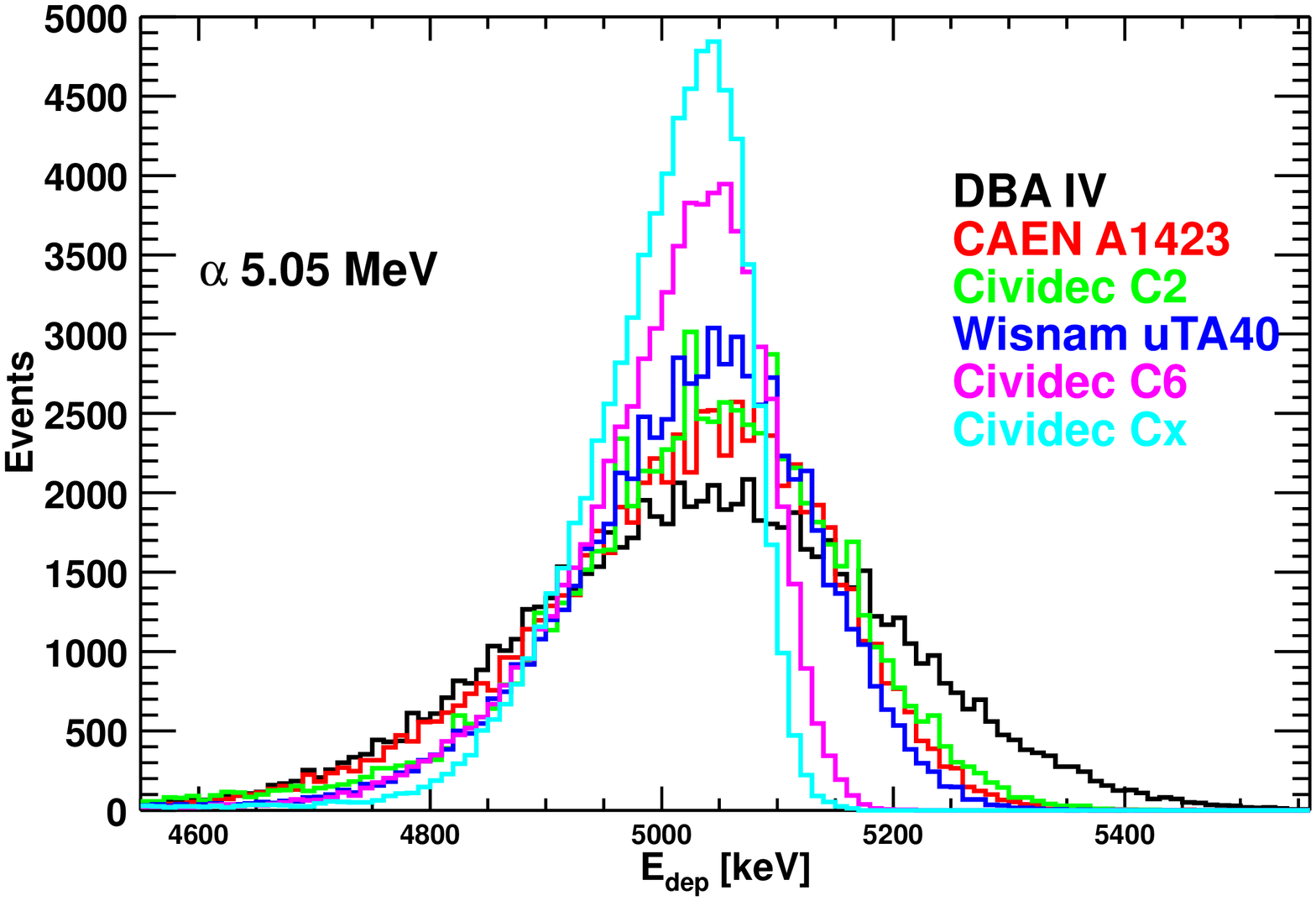}~~~%
\includegraphics[bb=4cm 4cm 20cm 27cm, angle=270, scale=0.4]{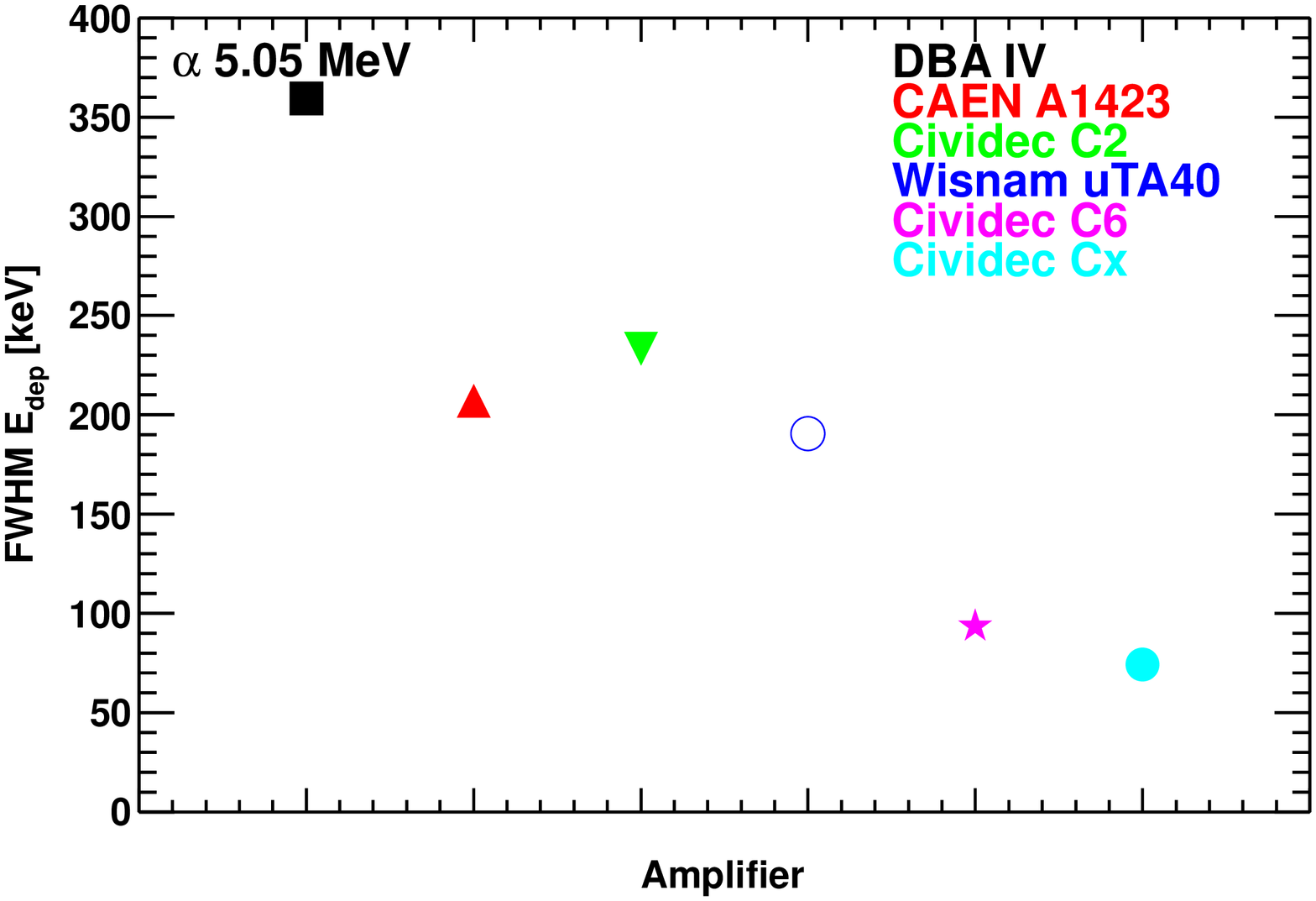}
\caption{\label{fig:edep_a5mev} Deposited energy distributions of $\alpha$s from $^{241}$Am source
after 6 mm of air for different amplifiers (left) and their FWHM resolutions (right).}
\end{center}
\end{figure}

The resolution on the measured charge generated by ionizing particle in CVD diamond
is dominated by the amplifier noise. Therefore to compare different amplifiers we measured
the charge generated by 5.5 MeV $\alpha$ particles produced by $^{241}$Am source
placed at 6 mm distance. To calibrate the energy scale we assumed that the peak
observed in the spectra corresponded to 5.05 MeV as shown in Fig.~\ref{fig:edep_a5mev}.

Current amplifiers have poorer resolution, which does not allow to distinguish
secondary $\alpha$ peaks at lower energy, leading to almost Gaussian peak shapes.
Instead both charge and transimpedance amplifiers show more detailed energy distributions.
The experimental resolution was measured for each amplifier by a fit of the observed peak by a Gaussian function.
For charge and transimpedance amplifiers only the high energy side of the peak was fitted.
The measured resolution of all current amplifiers lies around 200 keV FWHM,
except for DBA IV amplifier which shows a resolution inferior by a factor two.
The experimental resolution of charge and transimpedance amplifiers
was found to be larger than the expectation based on their noise.
This additional peak broadening is ascribed to fluctuations of $\alpha$ energy loss
in the 6 mm of air (about 400 keV).

\section{\label{sec:threshold}Minimal threshold}
\begin{figure}[!h]
\begin{center}
\includegraphics[bb=4cm 4cm 20cm 27cm, angle=270, scale=0.4]{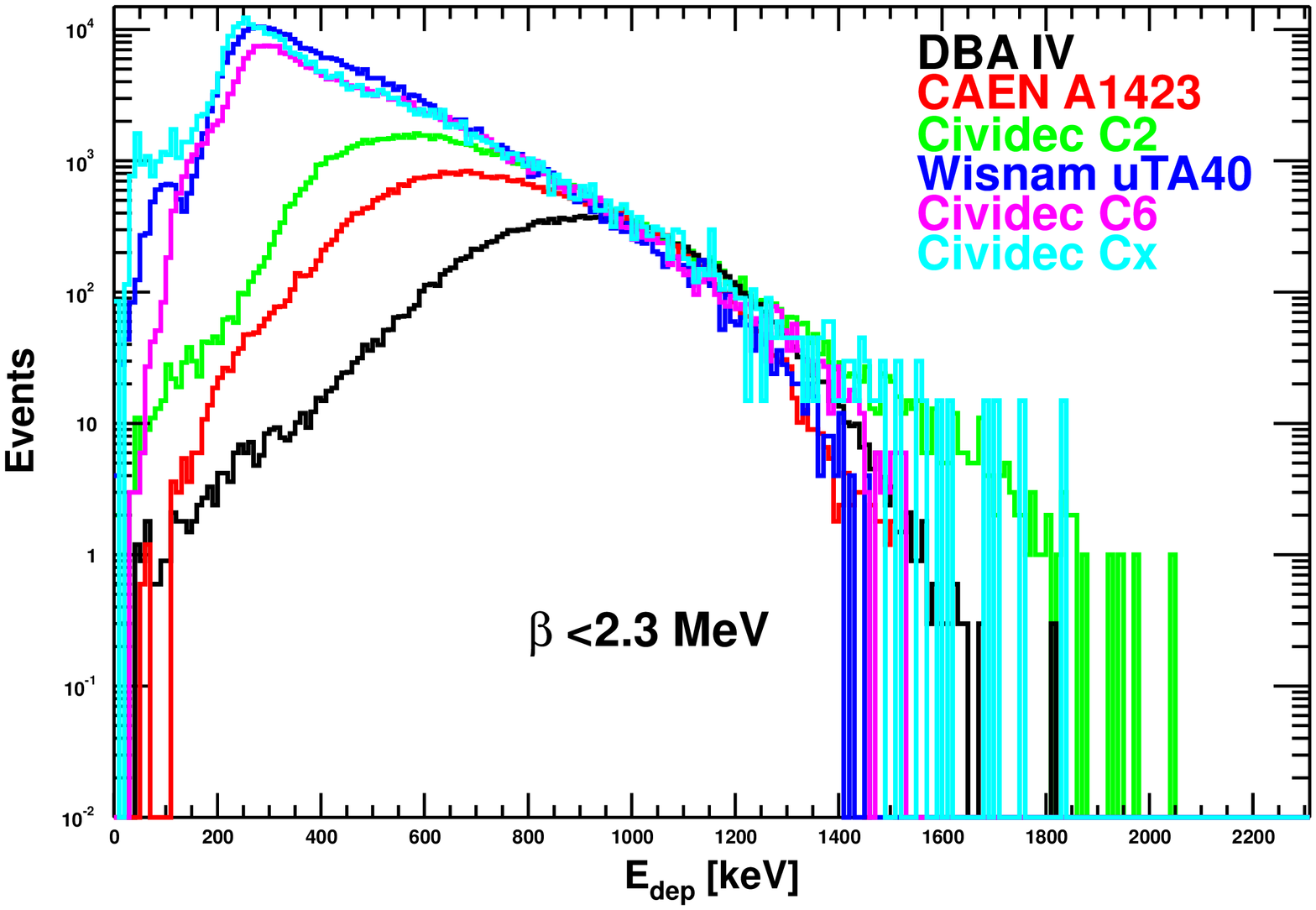}~~~%
\includegraphics[bb=4cm 4cm 20cm 27cm, angle=270, scale=0.4]{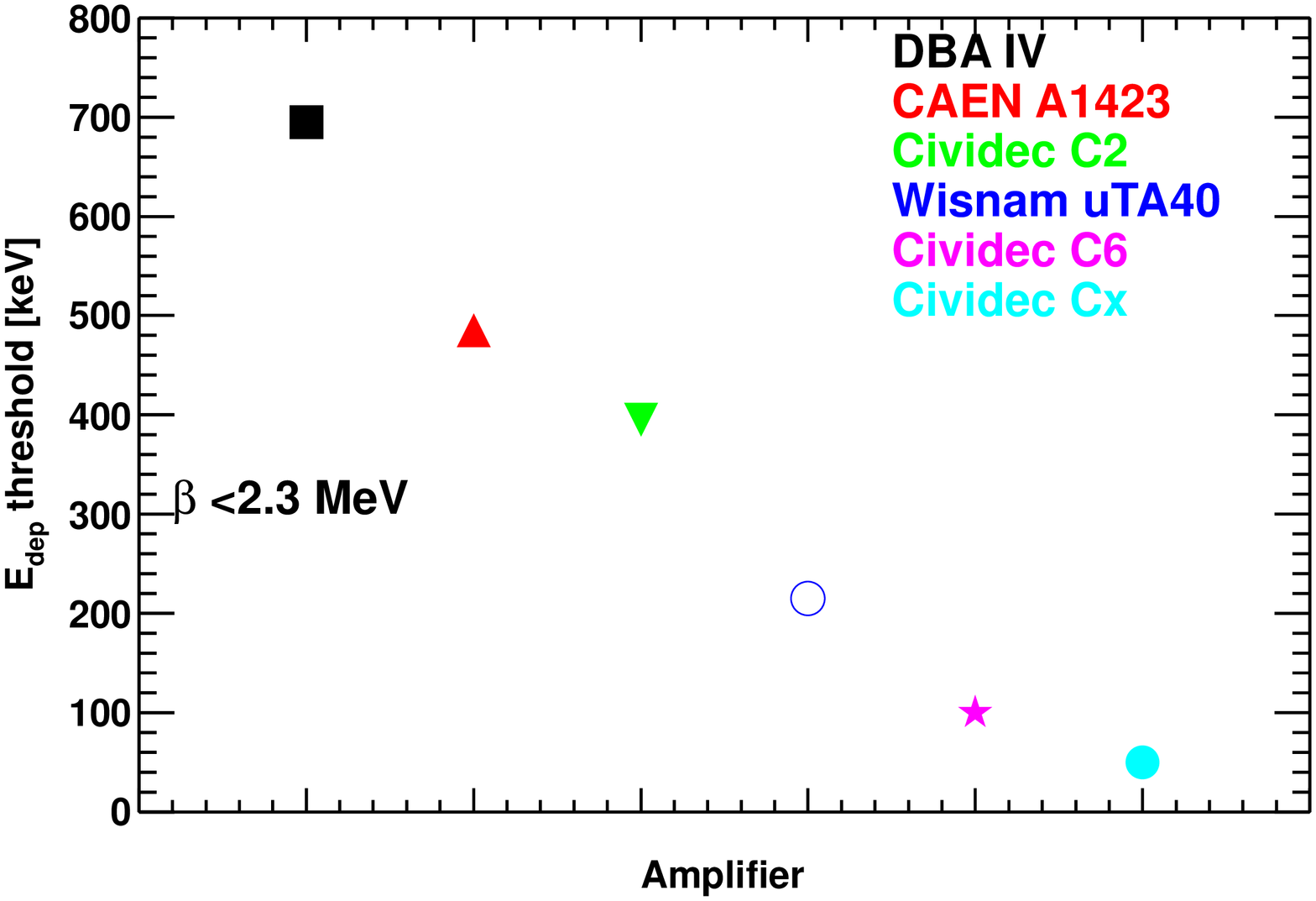}
\caption{\label{fig:edep_thr} Deposited energy distribution
measured in 500 $\mu$m CVD single crystal diamond detector
using $^{90}$Sr $\beta$ source with maximum energy of 2.3 MeV
acquired at the output of different amplifiers (left).
The threshold values obtained as the deposited energy value at half maximum height (right).}
\end{center}
\end{figure}

Another important aspect of diamond detector readout is the minimal achievable threshold.
This depends on the amplifier noise and gain characteristics. In order to measure
the threshold values we used $^{90}$Sr $\beta$ source with maximum $\beta$ energy of 2.3 MeV.
The digitizer threshold was set to the minimal value that limit noise rate to 100 Hz.
The response of the diamond detector, read out by means of different amplifiers,
is shown in the left plot of Fig.~\ref{fig:edep_thr}. For different amplifiers the $\beta$ decay spectrum
is cut at different energies. The cut-off energy determines the minimal achievable threshold.
In the right plot of Fig.~\ref{fig:edep_thr} the values of the threshold
estimated as the energy at which the spectrum reaches half of its maximum.

However, because the physical $\beta$-spectrum was not flat in energy,
at low deposited energies $<250$ keV these estimates were not always valid.
Indeed, the real threshold of the charge amplifier Cividec Cx was about 50 keV,
while the transimpedance amplifier Cividec C6 had the real threshold around 100 keV.
Instead, the threshold estimate for the fast current amplifier Wisnam $\mu$TA40
came out correctly to be 200 keV.

\section{\label{sec:time_res}Pulse shape and timing resolution}
\begin{figure}[!h]
\begin{center}
\includegraphics[bb=4cm 4cm 20cm 27cm, angle=270, scale=0.4]{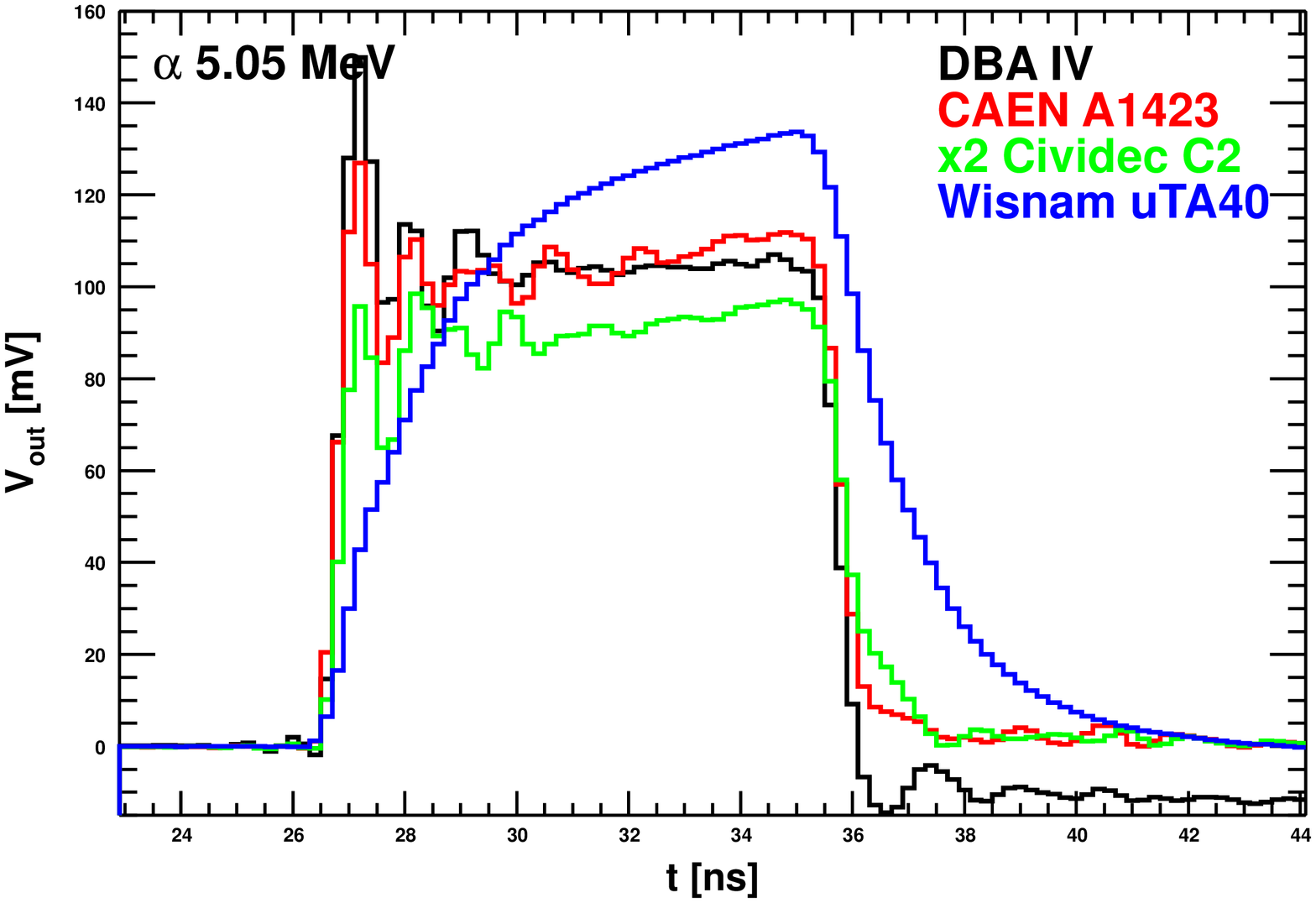}~~~%
\includegraphics[bb=4cm 4cm 20cm 27cm, angle=270, scale=0.4]{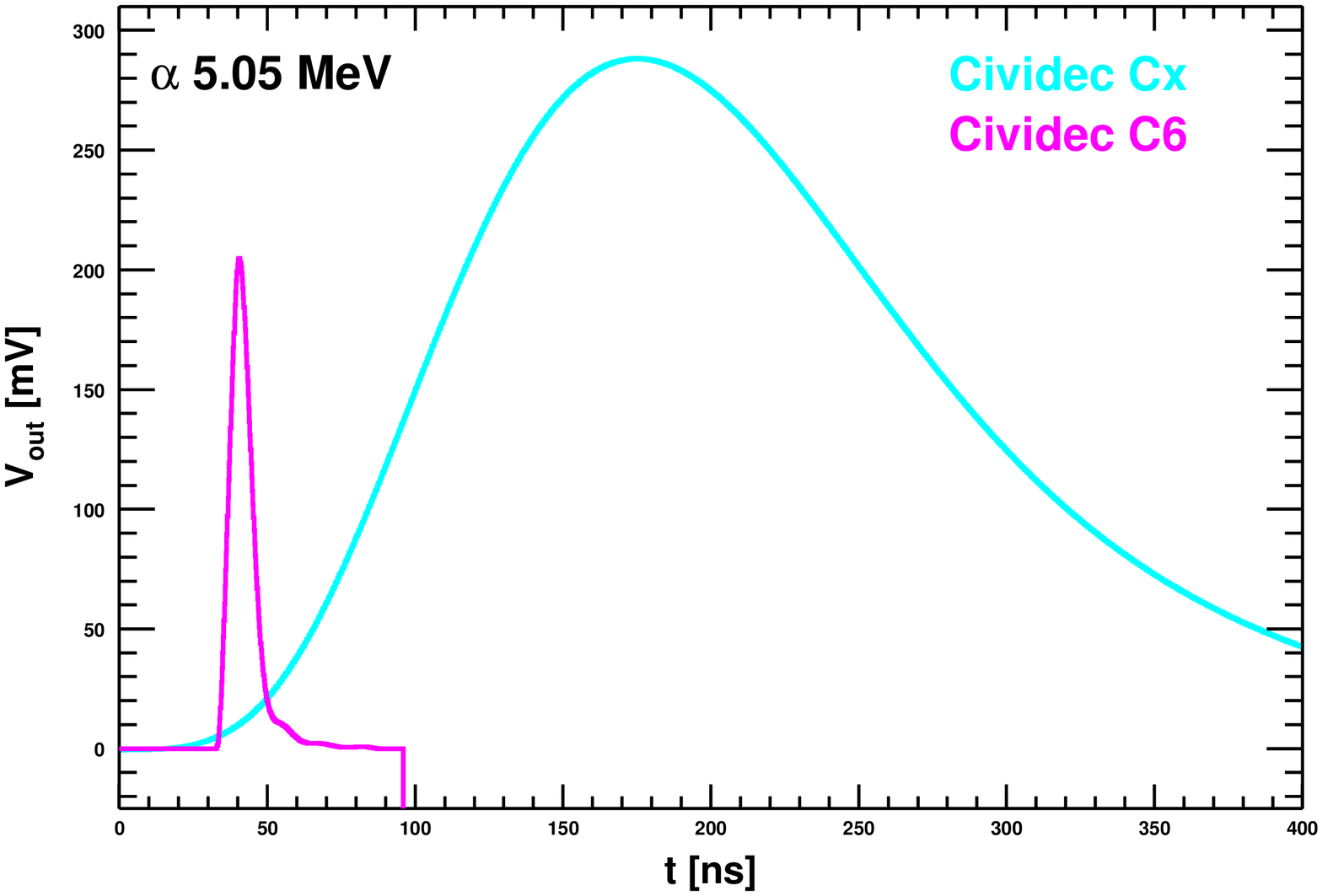}
\caption{\label{fig:pulse_shapes} Signal pulses due to $\alpha$s from $^{241}$Am source
in single crystal CVD diamond detector acquired at the output of different amplifiers.
Cividec C2 amplifier having lower gain than the others was rescaled by the factor of 2.}
\end{center}
\end{figure}

We performed timing and pulse shape measurements by studying coincidences between
signals from the two opposite electrodes of the diamond detector,
which were acquired by the SIS3305 digitizer in 5 Gs/s mode.
Waveforms of both signals were stored on disk.
The timing resolution was determined off-line.
First of all the signal shapes were acquired for each amplifier
as shown in Fig.~\ref{fig:pulse_shapes} in one-side readout mode.
The obtained pulse shapes were parametrized with opportune
RC and semi-Gaussian functions.
The pulse shapes of the three broadband amplifiers: DBA IV, CAEN A1423 and Cividec C2
are very similar and reproduce the corresponding input signals.
In these pulse shapes fast oscillations were observed
at the beginning of the pulse. The period of these oscillations is about 1 ns,
which is similar to twice the length of closed circuit from the amplifier input connector pole
through the diamond to the short at the opposite diamond electrode (about 16 cm).
Therefore, these oscillations can be explained by the reflections of the signal
due to incomplete impedance matching.
The Wisnam $\mu$TA40 amplifier showed reduced bandwidth of about 500 MHz,
compared to 1.5-2 GHz of DBA IV, CAEN A1423 and Cividec C2. The bandwidth was deduced from
its pulse shape, altered by the RC-integration and having much larger rise-time.
The transimpedance amplifier Cividec C6 gives relatively fast, 15 ns long, signals
with semi-Gaussian shaping rising in about 5 ns.
The charge amplifier Cividec Cx generates 350 ns long signals with
semi-Gaussian shaping and rise-time of about 130 ns. The latter signal
was not suitable for timing applications, but exhibited best energy resolution.
It was used only as a reference for diamond energy response.

\begin{figure}[!h]
\begin{center}
\includegraphics[bb=4cm 4cm 20cm 27cm, angle=270, scale=0.4]{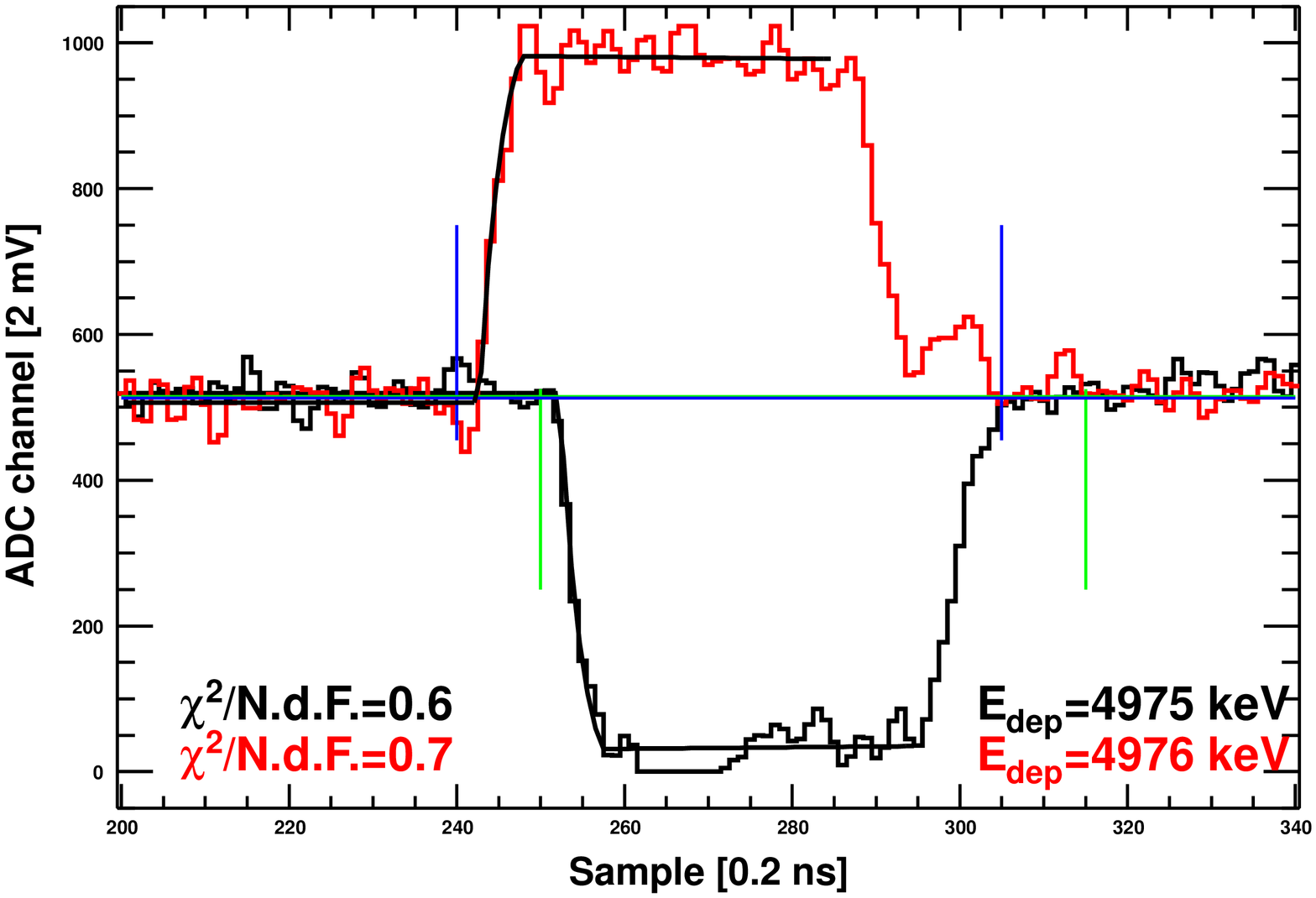}~~~%
\includegraphics[bb=4cm 4cm 20cm 27cm, angle=270, scale=0.4]{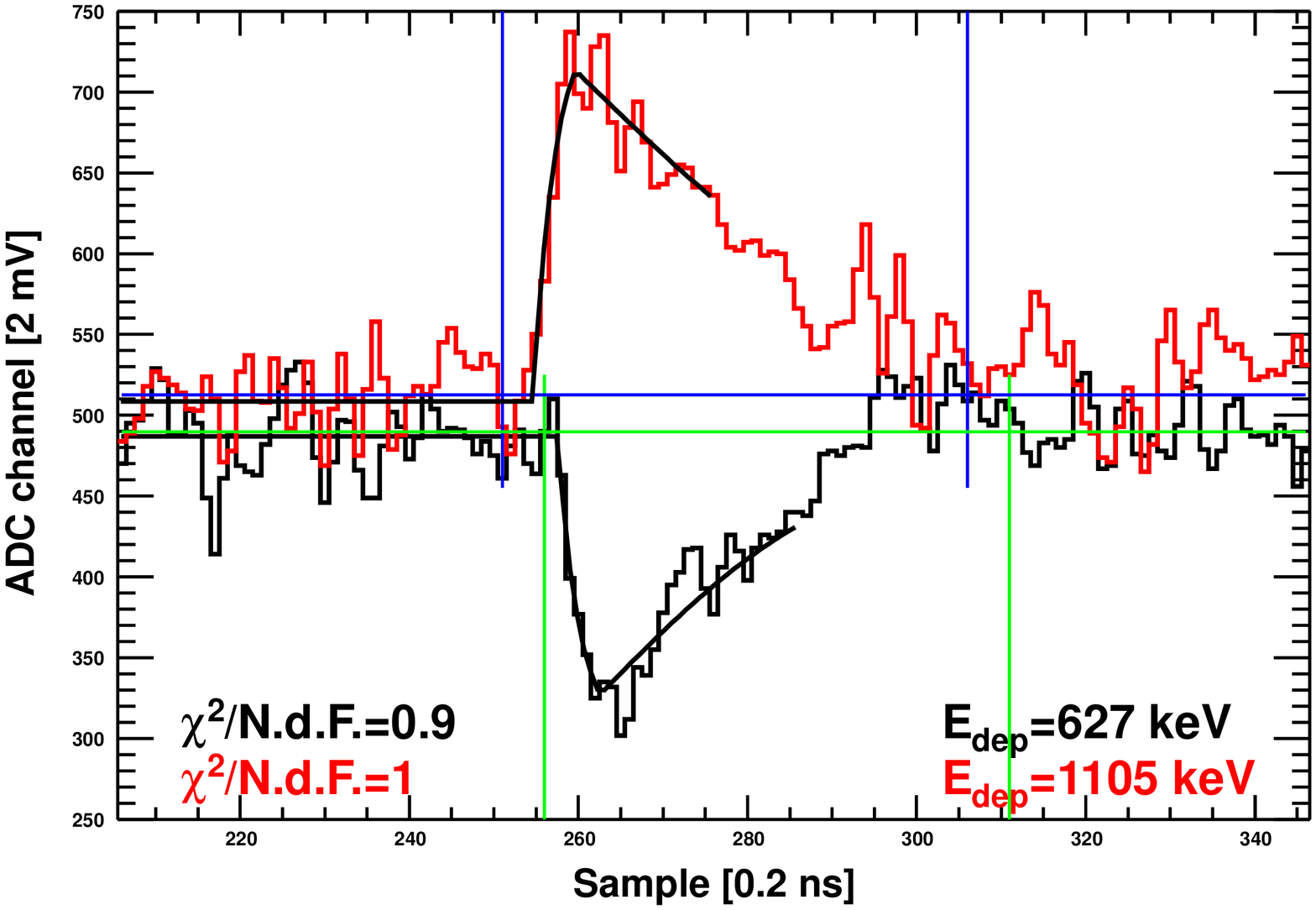}
\caption{\label{fig:coin_wf_c2} Pulses acquired in coincidence at two opposite contacts
of 500 $\mu$m CVD single crystal diamond detector
using 5 MeV $\alpha$ from $^{241}$Am source (left)
and $<$2.3 MeV $\beta$ $^{90}$Sr source (right).
The source was located at 6 mm distance from the diamond crystal in Air.}
\end{center}
\end{figure}

\begin{figure}[!h]
\begin{center}
\includegraphics[bb=4cm 4cm 20cm 27cm, angle=270, scale=0.4]{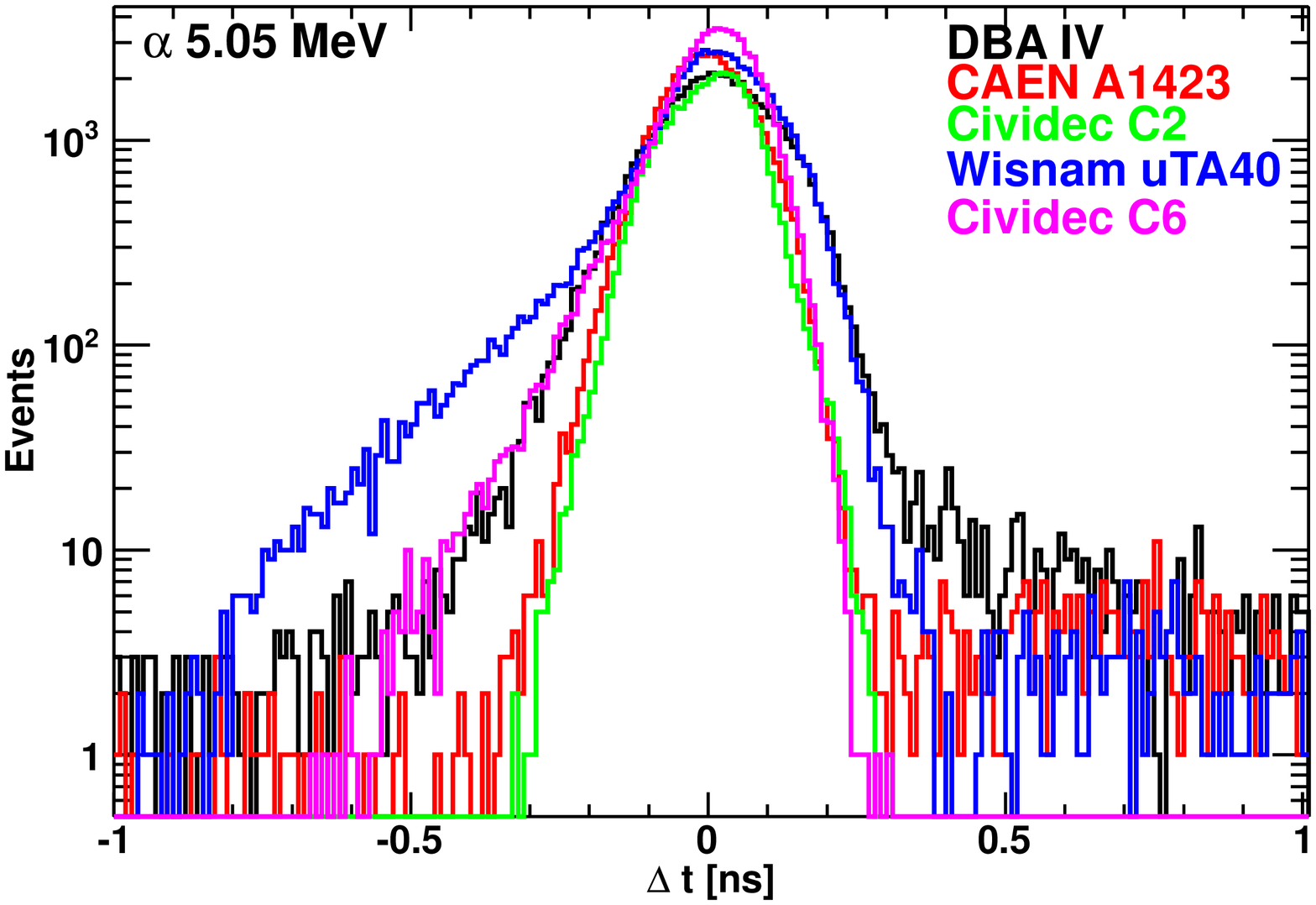}~~~%
\includegraphics[bb=4cm 4cm 20cm 27cm, angle=270, scale=0.4]{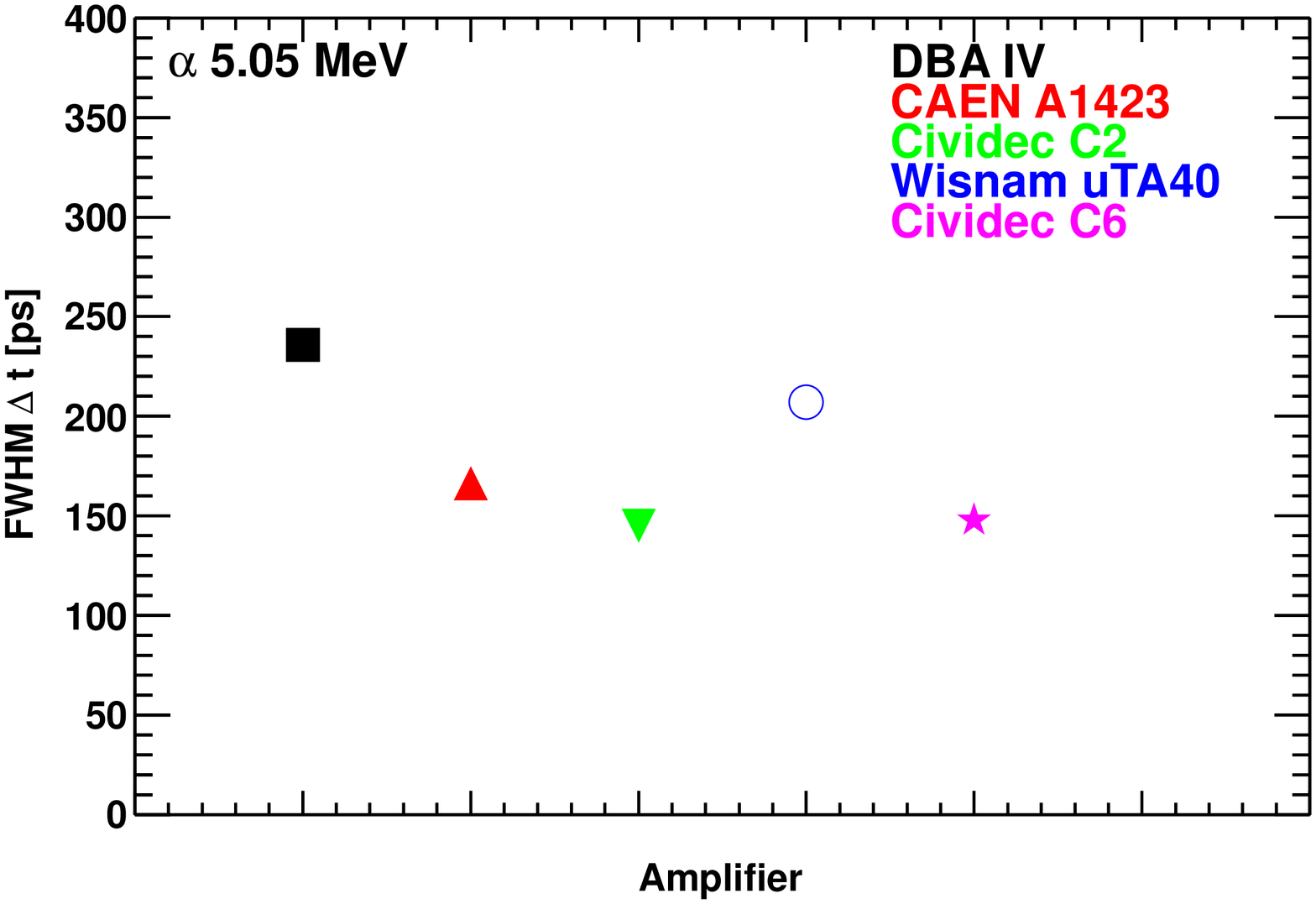}
\caption{\label{fig:dt_res} Timing difference between triggers from the two opposite contacts
of 500 $\mu$m CVD single crystal diamond detector (left)
and the corresponding timing resolution (right).
The data were obtained with 5 MeV $\alpha$s from $^{241}$Am source.}
\end{center}
\end{figure}

For each measured coincidence event both waveforms, shown for example in Fig.~\ref{fig:coin_wf_c2},
were fitted with selected functions to determine two start times. The difference between these two start times
was taken to be the coincidence time difference.
For large signals of 5.05 MeV $\alpha$s shown in Fig.~\ref{fig:dt_res}
the coincidence time difference is nearly Gaussian,
except for Wisnam $\mu$TA40 amplifier, which exhibits some tail at the l.h.s.
of the main peak. This tail is due to inaccurate description of the pulse shape.
The best overall timing resolution at 5.05 MeV was about 150 ps FWHM,
achieved with Cividec C2 and Cividec C6 amplifiers.
Wisnam $\mu$TA40 and DBA IV amplifiers showed $>200$ ps resolution at 5.05 MeV.

For smaller signals of 800 keV left by $\beta$s from $^{90}$Sr source the timing
resolution was considerably worse.
The energy dependence of the measured timing resolution can be compared to the naive expectation:
\begin{equation}
\sigma_t = \frac{\sigma_{noise}}{dS/dt|_{rise}} ~,
\end{equation}
\noindent which, assuming linear signal rise, can be rewritten as:
\begin{equation}\label{eq:dt_res}
\sigma_t \simeq \frac{t_{rise}}{S/N} ~,
\end{equation}
\noindent where signal-to-noise ratio $S/N$ can be calculated as the input signal peak voltage $V_{peak}$
to the input referred noise RMS $V_{noise}^{rms}$:
\begin{equation}
S/N=\frac{V_{peak}}{V_{noise}^{rms}} ~.
\end{equation}
\noindent Assuming triangular pulse shape, as in case of signals produced by $\beta$ particles, the peak voltage can be related
to the total energy deposited in the detector as following:
\begin{equation}
e \frac{E_{dep}}{E_{eh}}=\frac{1}{2} \frac{V_{peak}}{R_{in}} \Bigl[ t_{rise} +t_{fall} \Bigr] ~,
\end{equation}
\noindent where $R_{in}$ is the amplifier input impedance. The signal width, in turn,
can be expressed as following:
\begin{equation}
t_{rise} +t_{fall} = t_{len.} \simeq \frac{d}{v_{eh}(V_{bias})} ~,
\end{equation}
\noindent where carrier drift velocity $v_{eh}$ for a given bias voltage $V_{bias}$
can be measured as the length of $^{241}Am$ $\alpha$ signals:
\begin{equation}
v_{eh}(V_{bias})=\frac{d}{t_{len.}^{Am}} ~.
\end{equation}

In our case at $V_{bias}=1$ V/$\mu$m and $d=500$ $\mu$m we measured $t_{len.}^{Am}=9$ ns.
This corresponds to the carrier mean velocity:
\begin{equation}
v_{eh}(1 V/\mu m)=\frac{500\mu m}{9 ns} = 56 \frac{\mu m}{ns} = 2\times 10^{-4} c ~.
\end{equation}

Combining the above equations, the input signal peak voltage can be rewritten as following:
%\begin{equation}
%V_{peak} = 2e\frac{E_{dep}}{E_{eh}} \frac{R_{in}}{t_{rise} +t_{fall}} =
%2\times 1.6\times 10^{-4} fC \frac{E_{dep}}{13 eV} \frac{50\Omega}{10 ns} \frac{R_{in}/50\Omega}{(t_{rise} +t_{fall})/10 ns}
%\end{equation}
%
\begin{equation}
V_{peak} = 123 \mu V \frac{E_{dep}}{1 MeV} \frac{R_{in}/50\Omega}{(t_{rise} +t_{fall})/10 ns} ~.
\end{equation}

Finally substituting this into Eq.~\ref{eq:dt_res} we obtain the 
approximate expression for timing resolution:
\begin{equation}
\sigma_t \simeq t_{rise} \frac{V_{noise}^{rms}}{123 \mu V}
\frac{1 MeV}{E_{dep}} \frac{(t_{rise} +t_{fall})/10 ns}{R_{in}/50\Omega} ~.
\end{equation}

In the case of Cividec C2 amplifier $V_{noise}^{rms}\sim 20 \mu V$ and
therefore at $E_{dep}=0.8$ MeV we obtain:
\begin{equation}
\sigma_t \simeq 0.2 t_{rise}
\end{equation}
The above expression describes well the measured timing resolution for $t_{rise} \simeq 613$ ps.
%which corresponds to 4 200 ps digitizer samples.
This value is in agreement with observed rise time in Fig.~\ref{fig:coin_wf_c2},
%but factor of four larger than the expected $0.34/2$ GHz $=$170 ps.
and consists of detector capacitance $\tau_{in}=R C_{in}=50\Omega\times 6.6 pF=330$ ps,
Cividec C2 amplifier risetime of 170 ps and a smaller contribution due to
secondary amplifier and cables.

In the case of Cividec C6 amplifier the resolution can be estimated from its
sensitivity of 3 mV/fC and its linear gain $G\simeq 600$:
\begin{equation}
V_{peak} = \frac{36.9 mV}{G} \frac{E_{dep}}{1 MeV} ~.
\end{equation}
\noindent This combined with the amplifier noise value $V_{noise}^{rms}\sim 6 \mu V$
gives the timing resolution:
\begin{equation}
\sigma_t \simeq 0.1 t_{rise} \frac{1 MeV}{E_{dep}} ~.
\end{equation}
\noindent At $E_{dep}=0.8$ MeV and Cividec C6 rise time $t_{rise} \simeq 5$ ns
this estimate would imply timing resolution of $\sigma_t^{th} \simeq 625$ ps,
which is factor of 6 larger than the measured $\sigma_t^{exp} \simeq 105$ ps.
Our noise RMS was measured with 2 GHz bandwidth for a 100 MHz amplifier,
using instead the declared noise value $V_{noise}^{rms}\sim 1.2 \mu V$ allowed to describe the measured timing resolution.

\begin{figure}[!h]
\begin{center}
\includegraphics[bb=4cm 4cm 20cm 27cm, angle=270, scale=0.4]{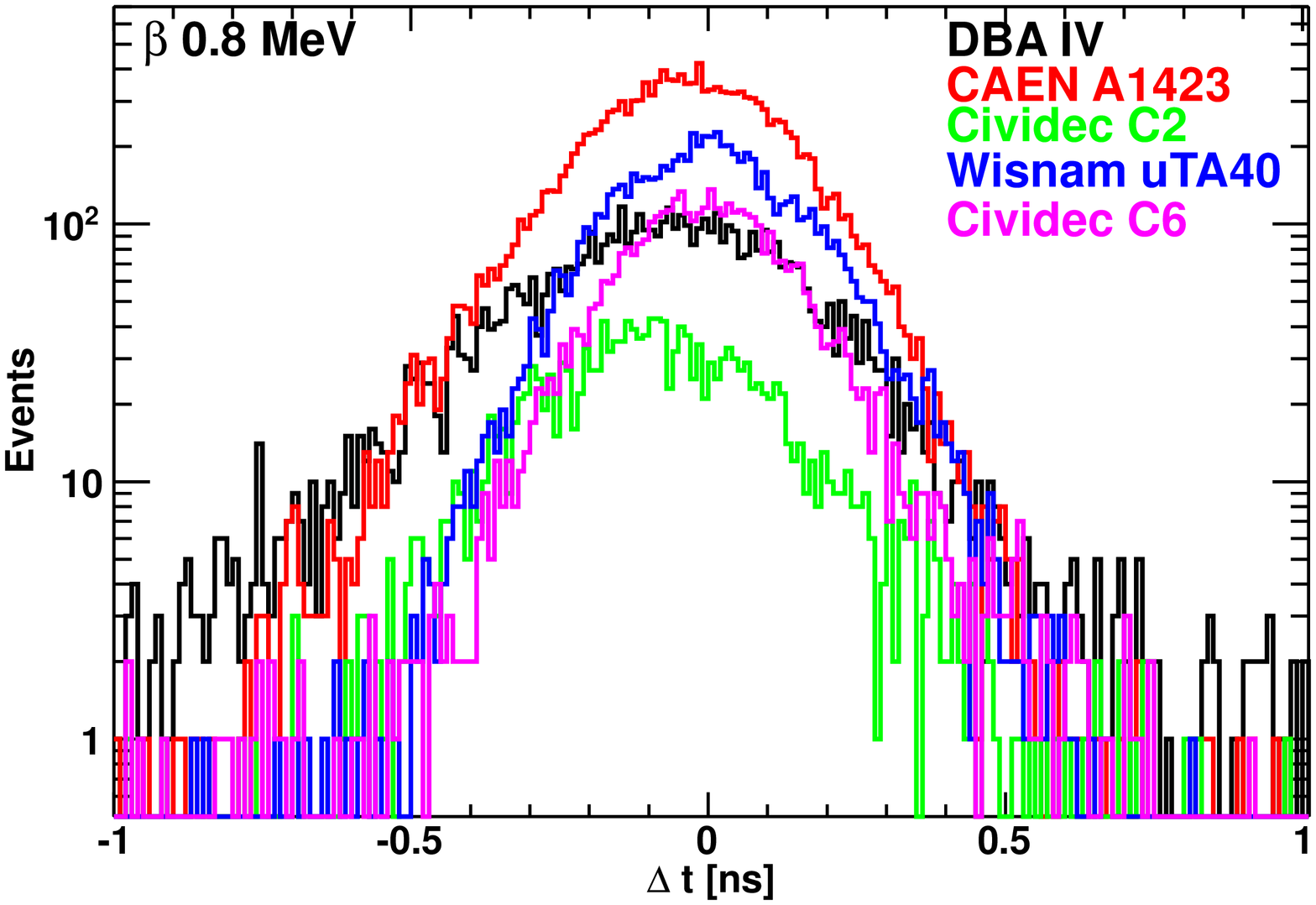}~~~%
\includegraphics[bb=4cm 4cm 20cm 27cm, angle=270, scale=0.4]{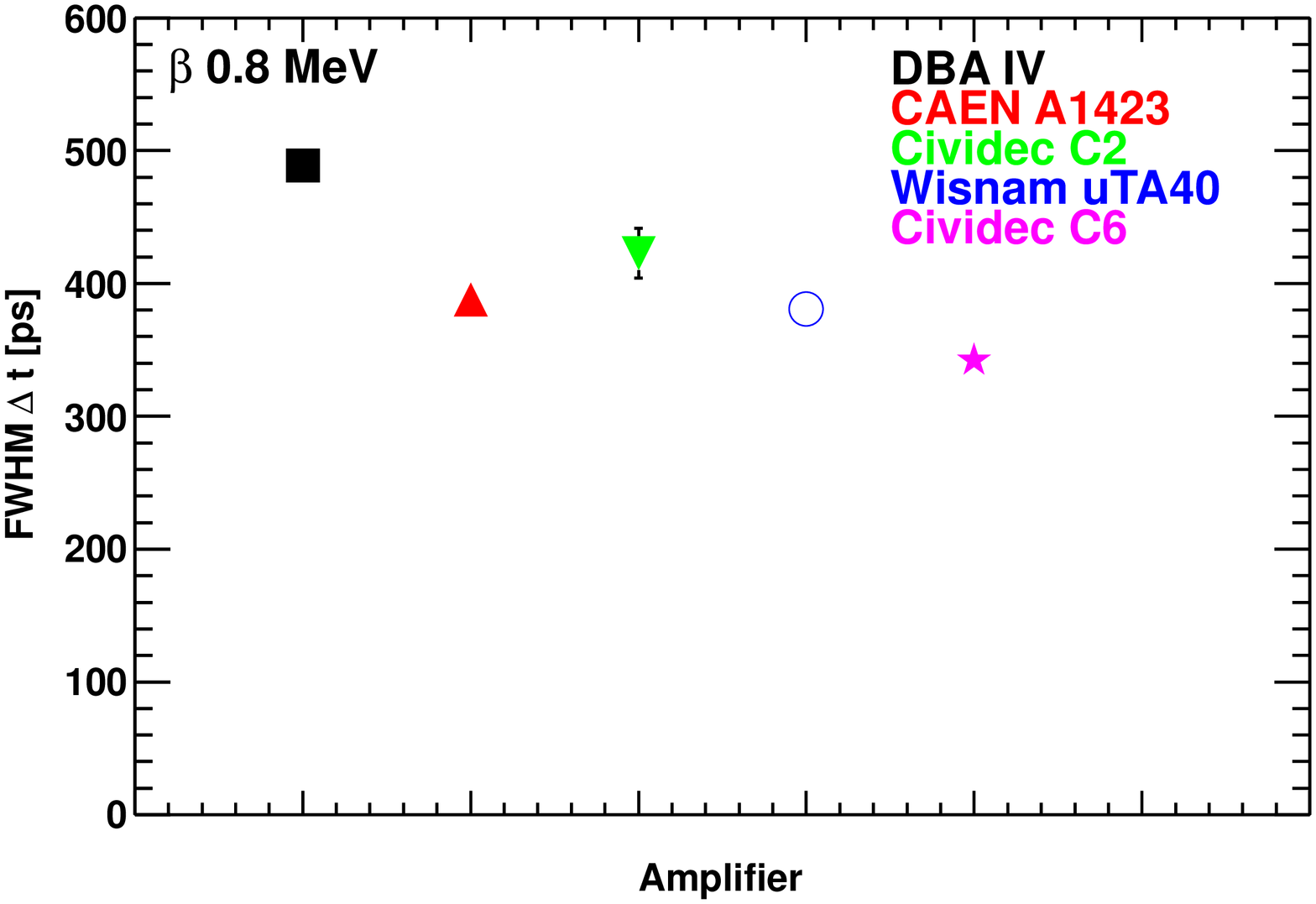}
\caption{\label{fig:dt_beta} Timing difference between triggers from the two opposite contacts
of 500 $\mu$m CVD single crystal diamond detector (left)
and the corresponding timing resolution (right).
The data were obtained with $<2.3$ MeV $\beta$ from $^{90}$Sr source
and selecting events whose deposited energy was from 0.7 to 0.9 MeV.}
\end{center}
\end{figure}

\begin{figure}[!h]
\begin{center}
\includegraphics[bb=4cm 4cm 20cm 27cm, angle=270, scale=0.4]{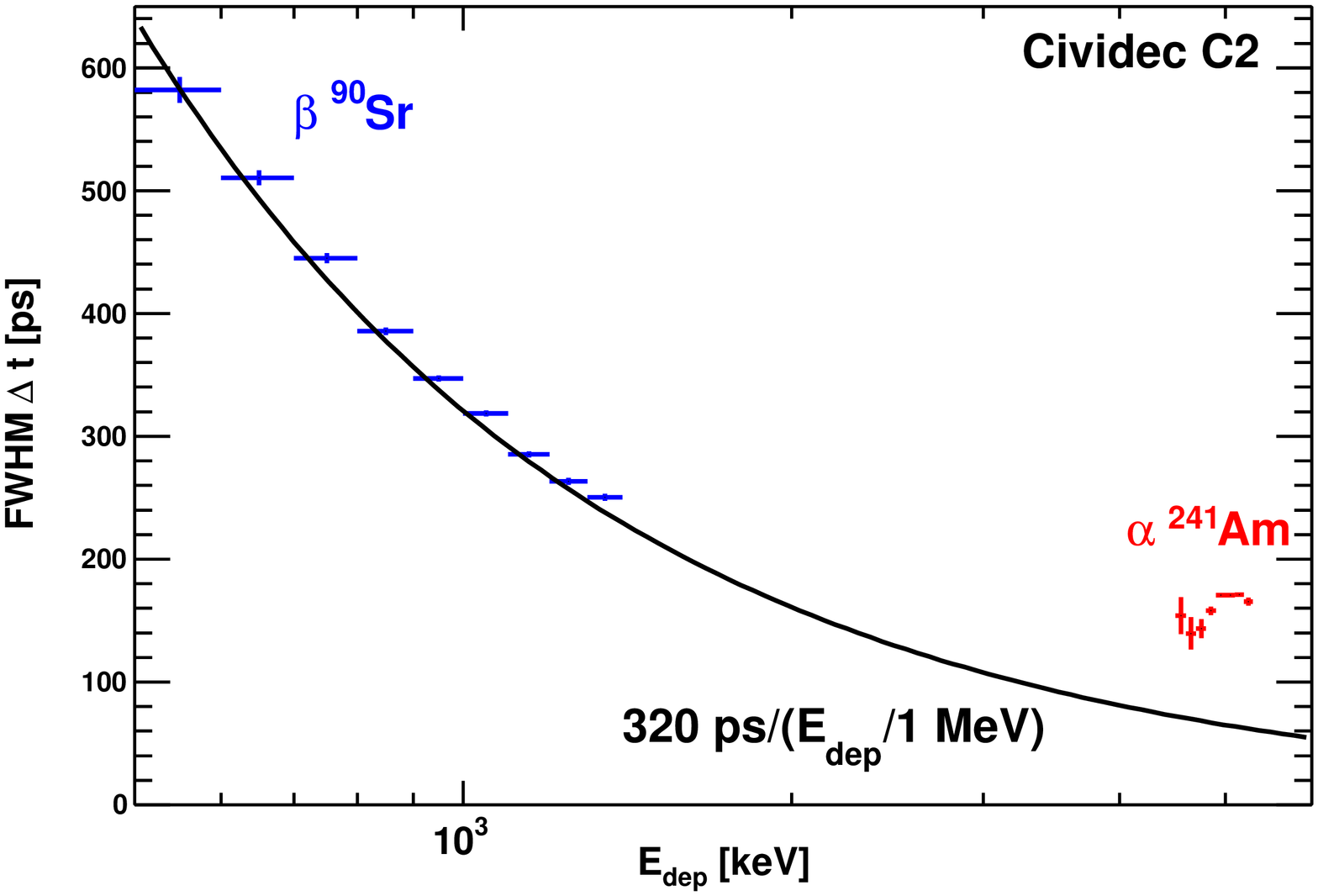}~~~%
\includegraphics[bb=4cm 4cm 20cm 27cm, angle=270, scale=0.4]{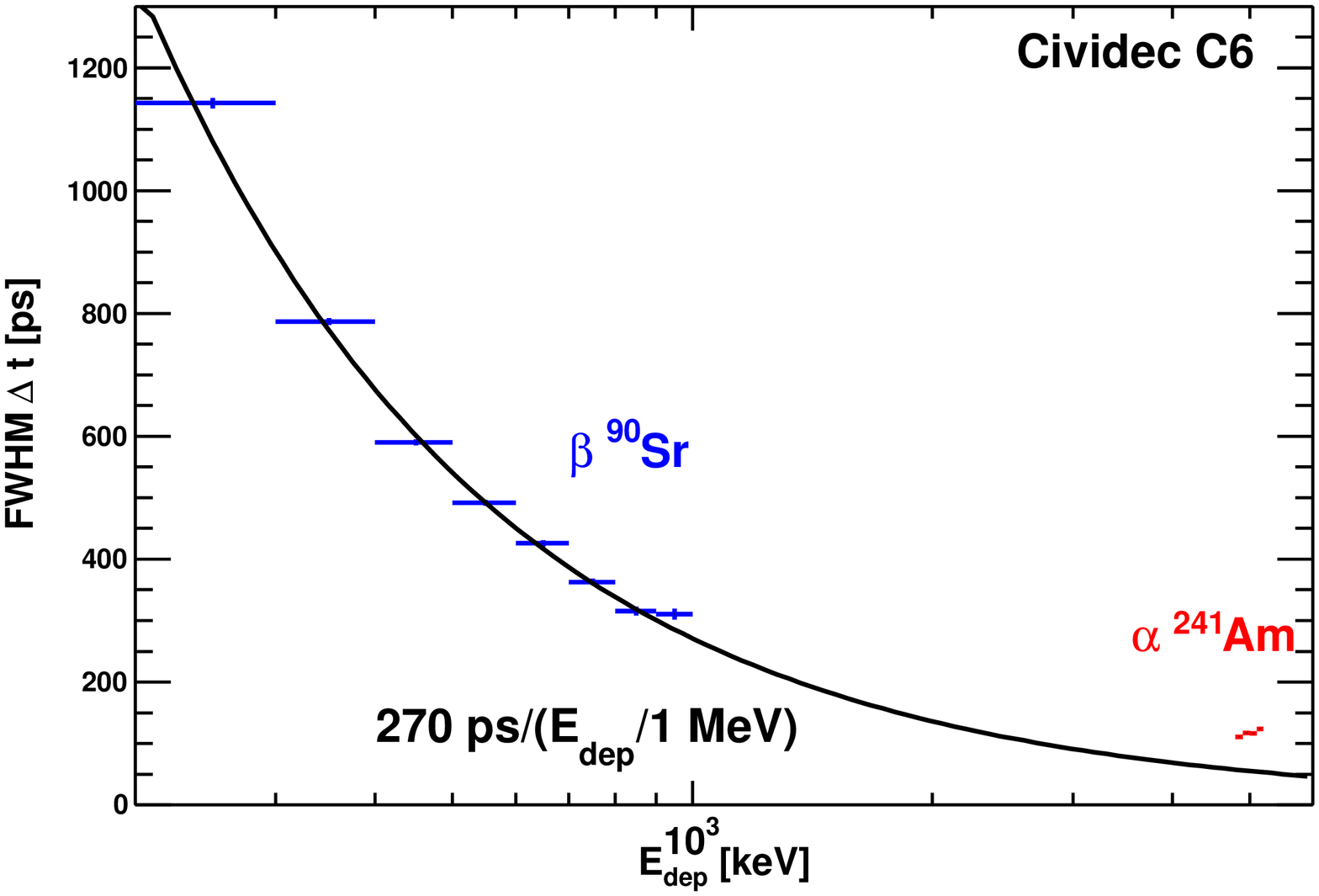}
\caption{\label{fig:dt_res_beta} Timing resolution of coincidence between two 500 $\mu$m CVD diamond signals
as a function of deposited energy
for Cividec C2 (left) and Cividec C6 (right) amplifiers.}
\end{center}
\end{figure}

\begin{table}[!h]
\begin{center}
\caption{\label{table:time_fwhm}Timing resolution FWHM at 1 MeV deposited energy
and standard deviation for MIP in 500 $\mu$m crystal (310 keV)
for different amplifiers. Starred values indicates that this amplifier cannot
reach 310 keV threshold and its resolution value is extrapolated.}
\vspace{2mm}
\begin{tabular}{|c|c|c|} \hline
Amplifier        & FWHM at 1 MeV  & $\sigma$ for MIPs in 500 $\mu$m \\
                 & [ps]           & [ps]    \\ \hline
DBA IV           & 377	          & 516*   \\ \hline
CAEN A1423       & 240	          & 329*   \\ \hline
Cividec C2       & 226	          & 310*   \\ \hline
Wisnam $\mu$TA40 & 181	          & 248   \\ \hline
Cividec C6       & 191	          & 262   \\ \hline
\end{tabular}
\end{center}
\end{table}

\section{\label{sec:cable_attenuation}Remote Detector}
In many experiments the amplifier cannot be installed close to the detector.
Insertion of a cable between detector and the amplifier introduces noise and
signal distortions, in particular when the impedance of the cable and amplifier is not well matched.
This is particularly important for charge and transimpedance amplifiers.
Standard broadband amplifiers featuring 50 $\Omega$ input impedance are almost
insensitive to the insertion of the input cable.
In our application the presence of a $>$1.5 m long cable is mandatory.
Hence we tested transimpedance and charge amplifiers Cividec C6 and Cividec Cx
with four different cables:
150 cm of 50 $\Omega$/87 pF/m SF105 (indicated as RG58),
150 cm of 75 $\Omega$/67 pF/m RG59,
166 cm of 93 $\Omega$/44 pF/m RG62,
616 cm of 185 $\Omega$/22.3 pF/m RG114.

\begin{figure}[!h]
\begin{center}
\includegraphics[bb=4cm 4cm 20cm 27cm, angle=270, scale=0.4]{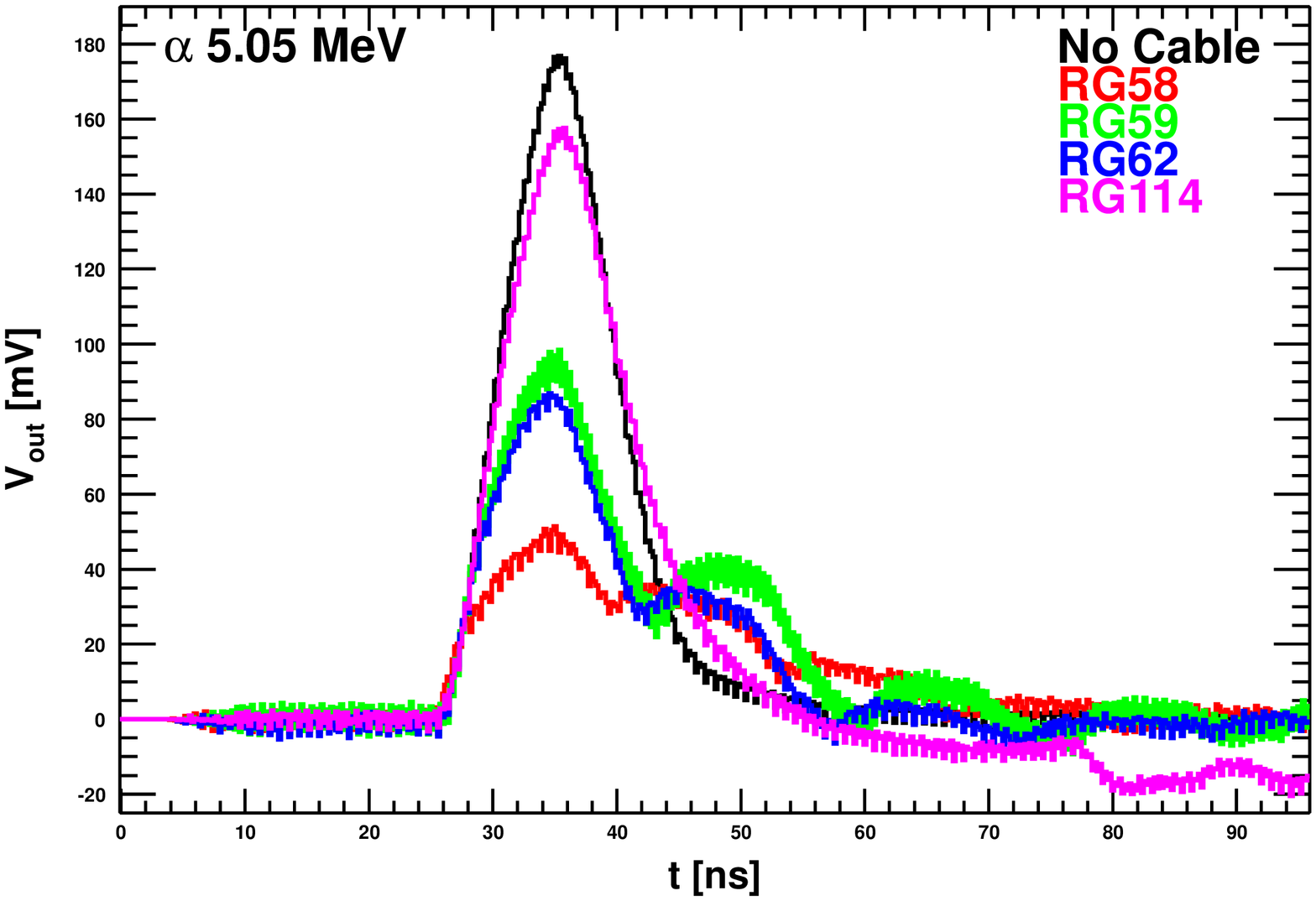}~~~%
\includegraphics[bb=4cm 4cm 20cm 27cm, angle=270, scale=0.4]{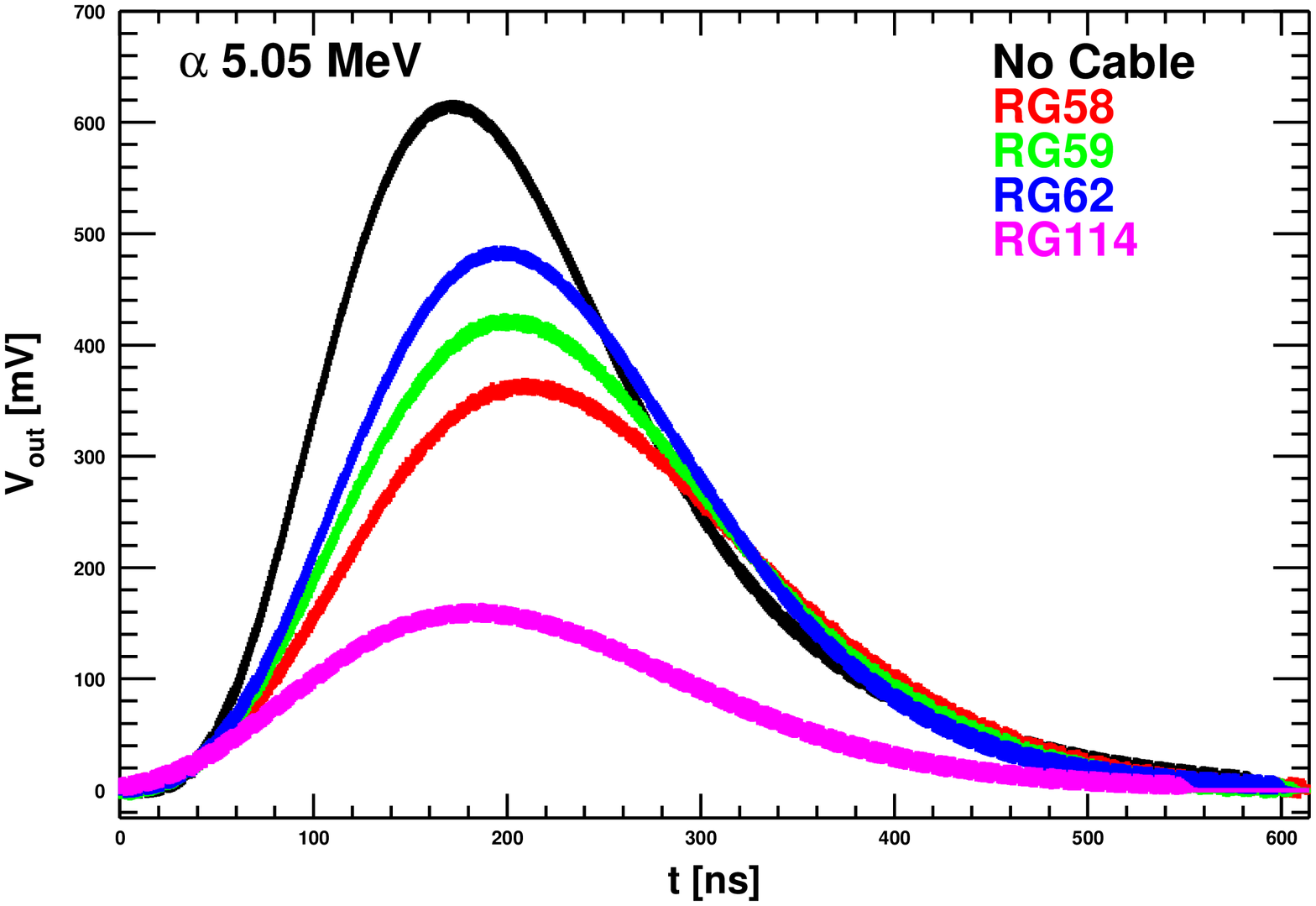}
\caption{\label{fig:cable_distortion} Distortion of the signal from diamond detector
introduced by insertion of various cables between detector and Cividec C6 transimpedance amplifier (left)
and Cividec Cx charge amplifier (right).}
\end{center}
\end{figure}

The comparison of signals produced by Cividec C6 and Cividec Cx amplifiers connected to the
detector via different cables is shown in Fig.~\ref{fig:cable_distortion}.

\begin{figure}[!h]
\begin{center}
\includegraphics[bb=4cm 4cm 20cm 27cm, angle=270, scale=0.4]{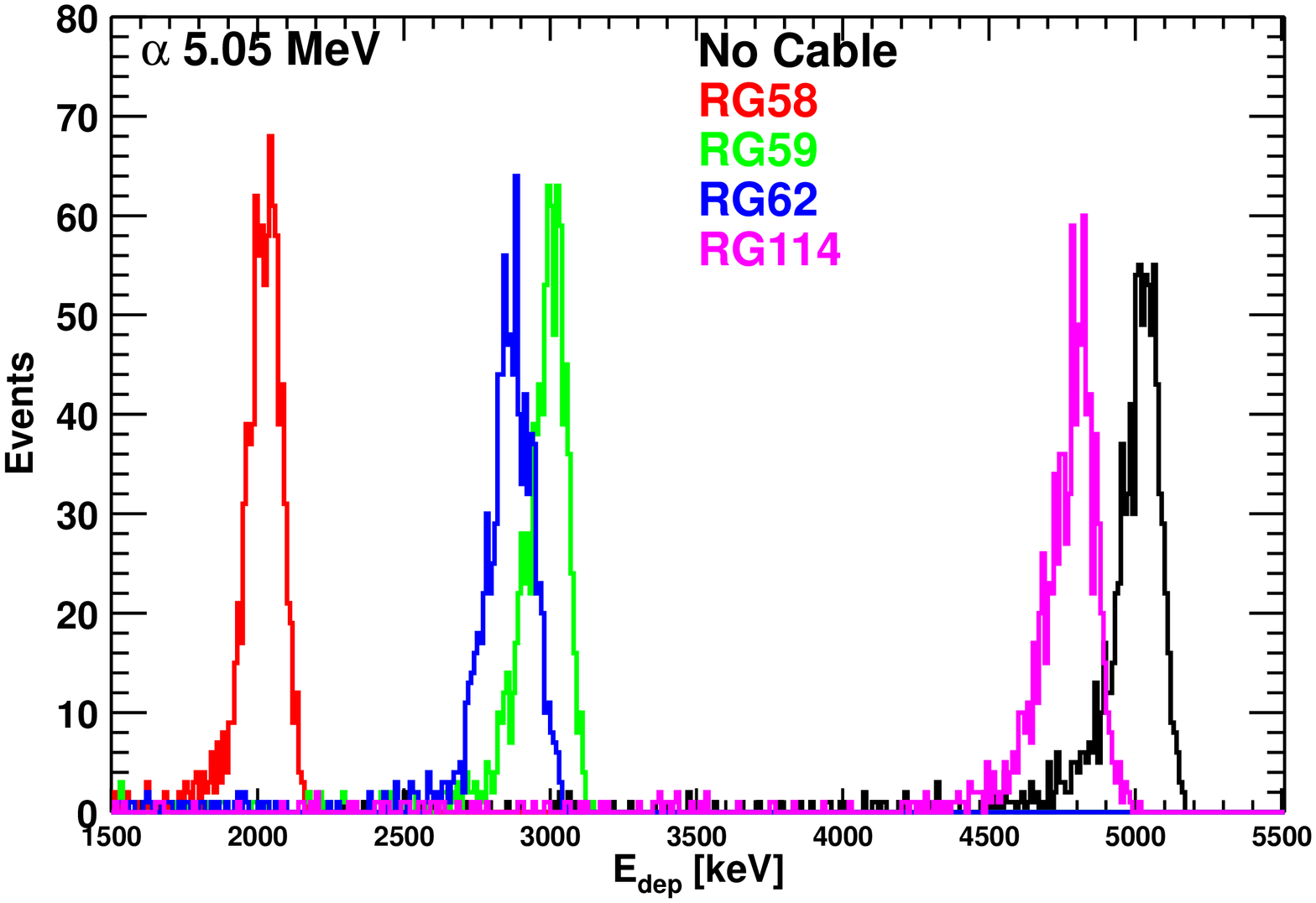}~~~%
\includegraphics[bb=4cm 4cm 20cm 27cm, angle=270, scale=0.4]{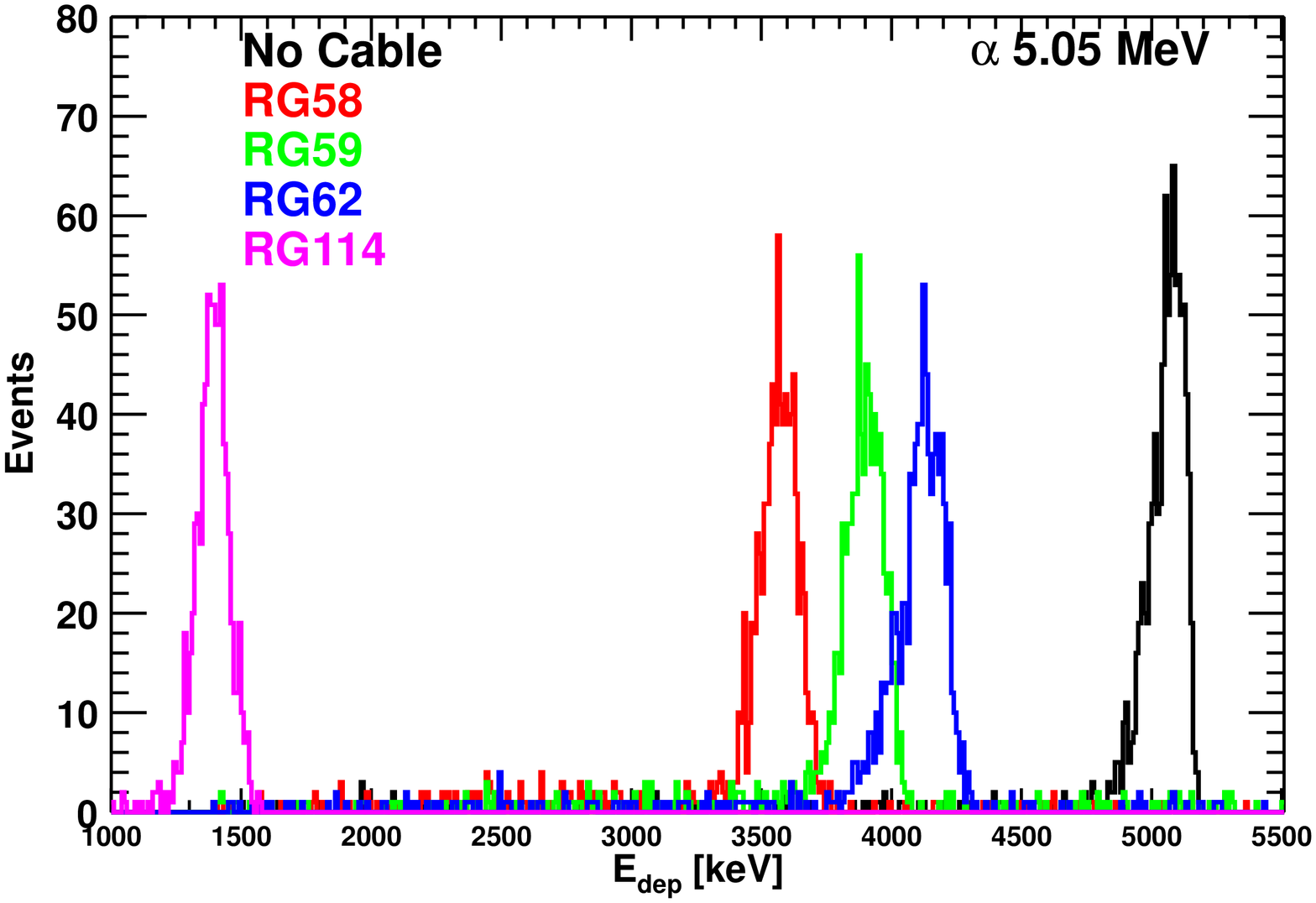}
\caption{\label{fig:cable_reduction} Reduction of the output signal amplitude
introduced by the insertion of various cables between the detector and Cividec C6 transimpedance amplifier (left)
and Cividec Cx charge amplifier (right).}
\end{center}
\end{figure}

Using the Cividec C6 amplifier, the insertion of a cable leads to strong signal reflections, reducing
for higher impedance cables and almost absent for 185 $\Omega$ RG114 cable\footnote{However, for another amplifier
of the same model we observed reflections also with RG114 cable.}.
From this we deduce that the amplifier input impedance is about 200 $\Omega$.
Using Cividec Cx amplifier cable insertion leads to signal suppression
proportional to the overall capacitance of the cable. Indeed, the RG62 cable with 44 pF/m capacitance
shows smallest suppression, while RG114 with 22.3 pF/m capacitance exhibits
larger suppression because its 4 times larger length.
This can be better quantified by measuring peak amplitude of the output signals
as shown in Fig.~\ref{fig:cable_reduction}.

\begin{figure}[!h]
\begin{center}
\includegraphics[bb=4cm 4cm 20cm 27cm, angle=270, scale=0.4]{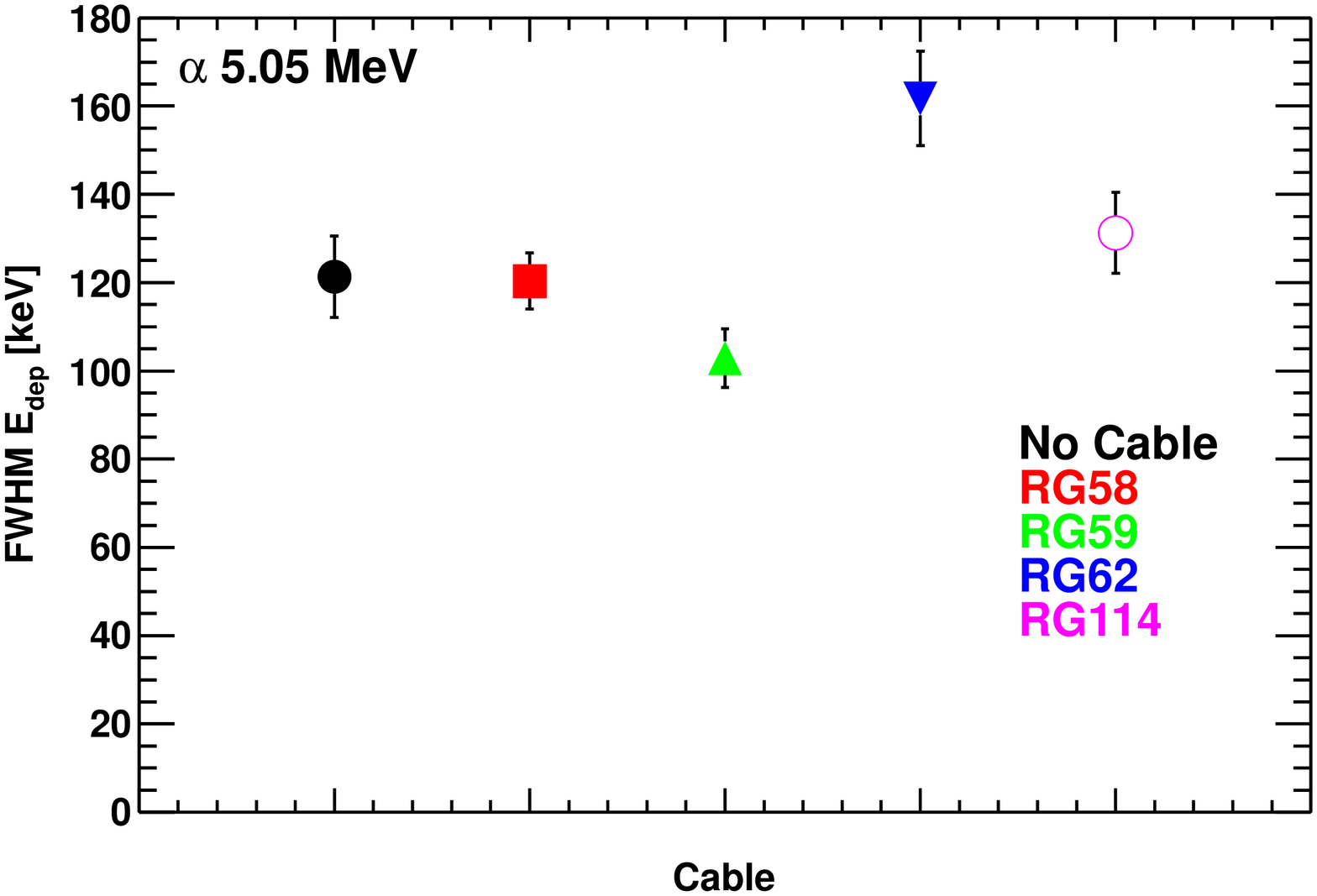}~~~%
\includegraphics[bb=4cm 4cm 20cm 27cm, angle=270, scale=0.4]{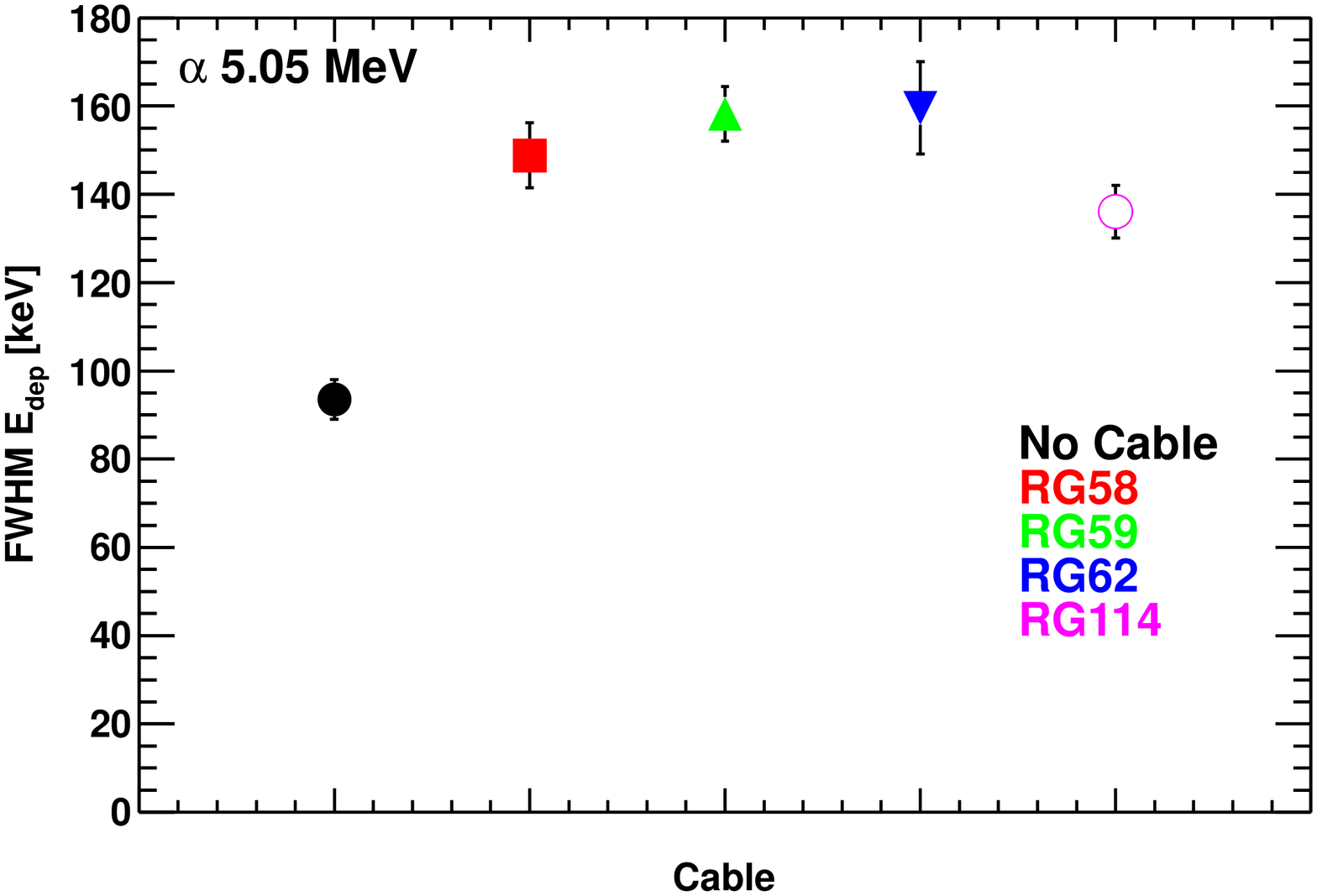}
\caption{\label{fig:cable_resolution} Deposited energy resolution variations
introduced by insertion of various cables between detector and Cividec C6 transimpedance amplifier (left)
and Cividec Cx charge amplifier (right).}
\end{center}
\end{figure}

The energy resolution variations due to insertion of different cables are different
for two amplifiers as shown in Fig.~\ref{fig:cable_resolution}.
Cividec C6 amplifier energy resolution changes by about 10\% with RG114 and RG62 cables,
while no effect is observed for other cables.
For Cividec Cx amplifier insertion of any cable leads to 30-40\% resolution loss.

\section{\label{sec:summary}Summary}
A number of modern commercial amplifiers for diamond detectors were characterized for typical nuclear and particle
physics applications. These include four broadband amplifiers one transimpedance amplifier and one charge amplifier.
We compared their energy and timing resolutions in order to select the best candidate for measuring
200 keV deposited energy signals.
Among these amplifiers only Cividec C6 and Cividec Cx amplifiers are able to reach
energy thresholds as low as 100 keV. However, the latter one is charge amplifier featuring
very long signals, not suitable for for timing application.
The obtained timing resolution at 200 keV of the only
amplifier which met our requirements, the Cividec C6, was found to be 1.2 ns (resolution of two Cividec C6 amplifiers in coincidence),
which was factor of 6 larger than our target resolution of 200 ps.
This resolution could be further reduced by a factor $\sqrt{2}$ by differential
readout of the detector, but would remain factor of 4 above the requirement.

The Wisnam $\mu$TA40 amplifier allows to measure 200 keV signals, but the resolution
is similar to the one of Cividec C6.

These results call for development of a new amplifier with signal-to-noise ratio
improved by a factor of 10 with respect to the tested ones.

%\newpage

\end{document}